\documentstyle[aps,eqsecnum]{revtex}

\begin{document}

 \draft

\title{\bf Effective field theory and the quark model}

\author{Loyal Durand\thanks{Electronic address: 
ldurand@theory2.physics.wisc.edu}, Phuoc Ha\thanks{Electronic address:
phuoc@theory1.physics.wisc.edu}, and Gregory Jaczko\thanks{Electronic
address:  Gregory.Jaczko@mail.house.gov}}
\address{ Department of Physics, University of Wisconsin-Madison,
Madison, WI 53706}

\date{\today}
\maketitle

\begin{abstract}

We analyze the connections between the quark model (QM) and the description of hadrons in the low-momentum limit of heavy-baryon effective field theory in QCD. By using a three-flavor-index representation for the effective baryon fields, we show that the ``nonrelativistic'' constituent QM for baryon masses and moments is completely equivalent through O($m_s$\/) to a parametrization of the relativistic field theory in a general spin--flavor basis. The flavor and spin variables can be identified with those of effective valence quarks. Conversely, the spin-flavor description clarifies the structure and dynamical interpretation of the chiral expansion in effective field theory, and provides a direct connection between the  field theory and the semirelativistic models for hadrons used in successful dynamical calculations. This allows dynamical information to be incorporated directly into the chiral expansion. We find, for example, that the striking success of the additive QM for baryon magnetic moments is a consequence of the relative smallness of the non-additive spin-dependent corrections.

\end{abstract}


\section{INTRODUCTION}
\label{sec:introduction}

The striking success of the nonrelativistic quark model (NRQM) in explaining the main features of baryon and meson masses and baryon magnetic moments suggests that its success is independent of the drastic approximations assumed in its typical formulations. To explore this point, we have examined the connection between the quark model (QM) approach to baryon masses and moments, and the rigorous relativistic effective field theory approach used in the chiral expansion of QCD. We show here that, as noted in \cite{DH_masses}, the QM for the static properties of baryons is simply  a parametrization of matrix elements in the underlying relativistic field theory in a general spin--flavor basis, where the flavor and spin variables can be identified with those of effective valence quarks. This identification holds exactly through first order in the chiral symmetry breaking mass variable $m_s$. The connection becomes clear when the chiral baryon fields are written in a natural form with three flavor and three spin indices. This change of basis clarifies the structure of the theory, including the origin of approximate SU(6) relations, and demonstrates the natural occurence of the QM spin-spin interaction terms in the baryon masses, and of effective quark moments in the description of the baryon magnetic moments \cite{phuoc-diss,jaczko-diss}.  

The change of basis also provides a direct connection between the effective field theory and semirelativistic models for hadrons \cite{brambilla} used in successful dynamical calculations \cite{kogut,isgur}, and allows dynamical information to be incorporated into the chiral expansion. We will show, in particular, that the three-flavor-index notation allows a natural classification of the spin-flavor correlations that appear in general matrix elements into those arising from effective one-, two-, and three-body operators. This classification corresponds directly to the underlying dynamics when spin-dependent forces and symmetry-breaking mass terms can be treated perturbatively, as seems to be the case. For example, the usual additive model for the baryon moments corresponds exactly to the one-body moment operator in the effective field theory. The model is successful because the non-additive two- and three-body operators arise from spin-dependent interactions and are correspondingly small.

The ``nonrelativistic'' aspects of the QM arise because the baryons are heavy, not because the dynamical quarks are heavy or nonrelativistic. The actual internal structure of the hadrons is absorbed into the momentum expansion of heavy-baryon chiral perturbation theory \cite{georgi-HBPT}, and the quark degrees of freedom move with the baryons. If these are sufficiently massive, the baryons may be treated as nonrelativistic with no recoil effects in loop diagrams, and the ``quark'' kinematics of the NRQM follow.
 
As we will show in detail in a subsequent paper \cite{DHJ}, the three-flavor-index representation of the fields allows an easy analysis of loop corrections \cite{phuoc-diss,jaczko-diss}, and shows why the residual loop corrections to the baryon masses \cite{DH_masses} and moments \cite{DH-loop-moments} are small. This is not the result of the smallness of individual loop corrections as such, but rather of the smallness of terms with new, nonadditive structures that violate the Gell-Mann--Okubo relations for masses and the Okubo relation for moments.

The paper is organized as follows. In Sec.\ \ref{sec:spin-flavor} we develop considerable backgound material on the three-flavor-index representation of the baryon fields, including calculational methods. In Sec.\ \ref{sec:chiral-interactions}, we rewrite the chiral expansion in this notation, comment on its connection with dynamics, and derive the baryon-meson couplings and the octet-decuplet mass splitting in the limit of equal quark masses.  We then analyze the baryon masses and moments to O($m_s$\/) in Secs. \ref{sec:masses} and \ref{sec:moments} using the new representation and its connection with dynamical models, and present concluding remarks in Sec.\ \ref{sec:conclusions}.

\section{EFFECTIVE FIELD THEORY IN A SPIN-FLAVOR BASIS}
\label{sec:spin-flavor}

\subsection{Heavy baryon chiral perturbation theory}
\label{subsec:heavy_baryons}

If the following sections, we will formulate heavy-baryon chiral perturbation theory (HBChPT) in a spin-flavor basis. This allows an easy connection of HBChPT to the underlying quark structure of the hadrons. We will then show that the results for the baryon masses and magnetic moments, taken at leading order in the chiral symmetry breaking quark mass matrix, are completely equivalent to those of the naive quark model. 

It will be useful as a first step to summarize the standard results we will need on the chiral expansion. This expansion is usually written in terms of matrix representations for the pseudoscalar meson and octet baryon operators,
\begin{equation}\label{pions}
\phi = {\textstyle \frac{1}{2}}\sum_{l=1}^8\,\lambda^l\phi^{\,l} = 
{\textstyle \frac{1}{ \!\!\sqrt{2}}}\left(
\begin{array}{ccc}
\frac{1}{\!\!\sqrt{2}}\,\pi^0 + \frac{1}{ \!\!\sqrt{6}}\,\eta & \pi^+ & K^+ \\ \pi^- & -\frac{1}{\!\!\sqrt{2}}\,\pi^0 + \frac{1}{ \!\!\sqrt{6}}\,\eta  & K^0 \\ K^- & \bar{K}^0 & - \frac{2}{\!\!\sqrt{6}}\,\eta
\end{array} \right) \,,
\end{equation}
\begin{equation}\label{baryons}
B = {\textstyle \frac{1}{\!\!\sqrt{2}}}\sum_{l=1}^8\,\lambda^l B^{\,l} = \left(
\begin{array}{ccc}
\frac{1}{\!\!\sqrt{2}}\,\Sigma^0 + \frac{1}{\!\!\sqrt{6}}\,\Lambda & \Sigma^+ & p \\ \Sigma^- & -\frac{1}{\!\!\sqrt{2}}\,\Sigma^0
 + \frac{1}{\!\!\sqrt{6}}\,\Lambda  & n \\-\Xi^- & \Xi^0 & -\frac{2}{\!\!\sqrt{6}}\,\Lambda
\end{array} \right) \,,
\end{equation}
where the $\lambda$\,s are the Gell-Mann matrices of SU(3). The matrices $\phi$ and $B$ transform on their first and second indices according to the {\bf 3} and $\bar{\bf 3}$ representations of SU(3) respectively, that is, as octet quark-antiquark ($q\bar{Q}$) combinations with
\begin{equation}
\phi \stackrel{U}{\longrightarrow} U\phi\,U^\dagger, \qquad B \stackrel{U}{\longrightarrow} UB\,U^\dagger.
\end{equation}
The $\phi$'s act as the Goldstone bosons of a broken ${\rm SU(3)}_L \otimes {\rm SU(3)}_R$ flavor symmetry \cite{Weinberg_Vol_2}.

Because the chiral expansion is a low-momentum expansion and the baryons are relatively massive, it is convenient in dealing with ChPT for baryonic processes to use the heavy-baryon formalism developed in Ref.\ \cite{georgi-HBPT} and extended to the chiral context in Ref.\  \cite{{Jen-HBChPT}}. This has been used to study a number of hadronic properties, for example, baryon masses \cite{Jen-HBChPT,Jen-masses,Bernard-masses,Bor-Mei-masses,Don-masses}, moments \cite{Jen-moments,Mei-moments,DH-ChPTmoments}, weak decays \cite{Jen-HBChPT,Jen-axial,Bos-L4}, and low-energy meson-baryon \cite{Jen-HBChPT} and electromagnetic \cite{BKKM-elec} interactions. The key ideas in heavy-baryon perturbation theory (HBPT) involve the replacement of the momentum $p^\mu$ of a nearly on-shell baryon by an on-shell momentum $m_0v^\mu$ plus a small additional momentum $k^\mu$, $p=m_0 v+k$, and the replacement of the baryon field operator $B(x)$ by an velocity-dependent operator $B_v(x)$ constructed to remove the dependence of the Dirac equation on the large momentum $m_0v^\mu$,
\begin{equation}
\label{B_v}
B_v (x) ={\textstyle\frac{1}{2}}(1+\not \!v) e^{i m_0 {\not v} v^{\mu} x_{\mu}}B(x), \quad v^\mu v_\mu=1\,. 
\end{equation}
The velocity-dependent perturbation expansion of the redefined theory involves modified Feynman rules and an expansion in powers of $k/m_0$ \cite{georgi-HBPT,Jen-HBChPT}. Here $m_0$ is an appropriate baryonic mass, $v^\mu$ is an on-shell four velocity, and it is assumed that $k\cdot v\ll m_0$. Velocity-dependent  Rarita-Schwinger decuplet fields $T_v^\mu$ can be defined in the same manner \cite{Jen-HBChPT}. We will work in the heavy baryon limit throughout the paper, and will henceforth drop the subscript $v$ on the fields $B_v$, $T_v$.

For later reference, we give the flavor-symmetric chiral Lagrangian for the modified fields at leading order in the momentum expansion, retaining all chiral invariants in the baryon fields with at most one derivative \cite{Jen-HBChPT}: 
\renewcommand{\arraystretch}{1.5}
\begin{eqnarray}\label{lag}
{\cal L}_0 &=& i\, {\rm Tr}\, \bar{B}\,(v\cdot {\cal D})B - \delta m_B {\rm Tr}\, \bar{B}B + 2D\, {\rm Tr}\, \bar{B}\, S^\mu \{ A_\mu, B \}
+ 2F\, {\rm Tr}\,  \bar{B}\, S^\mu\, [A_\mu, B]
\nonumber \\
&&-i\, \bar{T}^{\mu}\, (v \cdot {\cal D}) \,  T_\mu
+ \delta m_T \, \bar{T}^\mu\, T_\mu
+ {\cal C}\, (\bar{T}^\mu A_\mu B + \bar{B} A_\mu T^\mu) \\
&& +\ 2{\cal H}\,\bar{T}^\mu\, S_\nu\, A^\nu\, T_\mu
+ {\textstyle\frac{1}{4}}f^2\, \rm{Tr}\,\partial_\mu\Sigma \partial^\mu\Sigma \,. \nonumber
\end{eqnarray}
\renewcommand{\arraystretch}{1}
Here $\delta m_B = m_B-m_0$ and $\delta m_T = m_T-m_0$, ${\cal D}_\mu= \partial_\mu+[V_\mu,\, \cdot\, ]$ is the covariant chiral derivative, $S^{\mu}$ is the spin operator defined in \cite{Jen-HBChPT}, and $D$, $F$, ${\cal C}$, and ${\cal H}$ are strong interaction coupling constants. The vector and axial vector currents are given by
\begin{equation}
\label{Vcurrent}
V_\mu = {\textstyle \frac{1}{2}}\left(\xi \partial_\mu \xi^\dagger + \xi^\dagger\partial_\mu \xi\right) =  f^{-2} \left(\phi\partial_\mu\phi - \partial_\mu\phi \phi\right)+\cdots  \,, 
\end{equation}
\begin{equation}
\label{Acurrent}
A_\mu ={\textstyle \frac{i}{2}}\left(\xi \partial_\mu \xi^\dagger - \xi^\dagger\partial_\mu \xi\right) =  f^{-1}\partial_\mu \phi+\cdots \,,
\end{equation}
where
\begin{equation}
\label{xi}
\xi=e^{i\phi/f}, \quad \Sigma=e^{2i\phi/f}=\xi^{\,2},
\end{equation}
and $f \approx 93$ MeV is the pion decay constant. The currents are traceless,
\begin{equation}
\label{Tr=0}
{\rm Tr}\,V_\mu = {\rm Tr}\, A_\mu = 0,
\end{equation}
a condition that eliminates the potential invariants $({\rm Tr}\,\bar{B} S^\mu B )\,({\rm Tr}\,A_\mu)$ and $({\rm Tr}\,\bar{T}^\mu S^\nu T_\mu)\,({\rm Tr}\, A_\nu)$.

\subsection{Effective baryon and meson fields in a spin-flavor basis} 
\label{subsec:baryon_meson_fields}

\subsubsection{The octet-baryon fields}
\label{subsubsec:octet_3index}

The two-index matrix representation of the octet baryons  in Eq.\ (\ref{baryons}) hides their three-quark structure. This makes it difficult to trace the flow of flavor through a process involving baryons, or to connect the chiral picture with the underlying quark picture. We will therefore adopt a three-index description of the flavor structure of the baryons. This appears at the outset to be more complicated, but a three-flavor-index notation is already standard for the decuplet baryons.  The change in the description will allow a simple, detailed analysis of the structure of loop connections in a subsequent paper \cite{DHJ}. 

We consider, in particular, representations of the baryons constructed using the ``quark'' field $q_i^{\,\alpha a}$ and its conjugate $\bar{q}_i^{\,\alpha a}$ to carry the flavor, spin, and color structure of the baryons. Here $i\in u,\,d,\,s$ is the flavor index, and $a\in 1,\,2,\,3$ is the color index, and $\alpha$ is a Dirac spinor index. These fields transform under the vector or diagonal subgroup SU(3)$_V$ of the chiral ${\rm SU(3)}_L \otimes {\rm SU(3)}_R$ as fundamental and anti-fundamental representations, respectively
\begin{equation}
	q_i \stackrel{U}{\longrightarrow} U_{ii'} q_{i'} \quad {\rm and} \quad
	\bar{q}_i \stackrel{U}{\longrightarrow} U^*_{ii'}\bar{q}_{i'} = \bar{q}_{i'} U^\dagger_{i'i}\,.
\label{eq:Qtrans}
\end{equation}

The octet baryons are easy to represent in a three-flavor-index notation even though they involve flavor and spin combinations of mixed symmetry in an SU(3)$_f \otimes$ SU(2)$_{spin}$ decomposition. The key observation is that there is only one color-singlet combination of three anticommuting quark fields with total spin 1/2. This corresponds to a flavor octet. There is no flavor singlet.\footnote{The flavor-singlet and color-singlet components of $\bf 3 \otimes 3 \otimes 3$ are completely antisymmetric, giving overall symmetry. There is no completely antisymmetric combination of three spins in $\bf 2 \otimes 2 \otimes 2$, hence no $j=1/2$ flavor- and color-singlet combination of three quarks with the required overall antisymmetry.} The fermionic symmetry is built in automatically in a quark-field description. We can therefore determine the properties of the octet baryon fields trivially by combining two quark fields in a singlet spin state, multiplying by a third quark field which carries the total spin, and combining the color indices in a color singlet. The result is an octet field $B_{ijk}^{\,\gamma}$ 
\begin{equation}
\label{Bijk}
B_{ijk}^{\,\gamma} = {\textstyle\frac{1}{6}}\,\epsilon_{abc}\,q_i^{\,\alpha a}q_j^{\,\beta\, b}q_k^{\,\gamma\, c}\,(C\gamma^5)_{\alpha \beta}\,,
\end{equation}
where we have used the charge conjugation matrix $C=i\gamma^2\gamma^0$ to write $B_{ijk}^{\,\gamma}$ as a spinor product. $B_{ijk}^{\,\gamma}$ transforms under SU(3)$_V$ as
\begin{equation}
\label{baryon_transform}
B_{ijk}^{\,\gamma} \stackrel{U}{\longrightarrow} U_{ii'}U_{jj'}U_{kk'} B_{i'j'k'}^{\,\gamma}.
\end{equation}

As long as we are dealing with processes that do not involve color dynamics such as gluon emission or exchange, we can suppress the color factor $(1/\sqrt{6})\epsilon_{abc}$ and the color indices on the quark fields and treat the $q$'s as commuting rather than anticommuting fields, with
\begin{equation}
\label{Bnocolor}
B_{ijk}^{\,\gamma} \equiv {\textstyle\frac{1}{\!\!\sqrt6}}\,q_i^{\,\alpha}q_j^{\,\beta}q_k^{\,\gamma}\, (C\gamma^5)_{\alpha \beta}  = {\textstyle\frac{1}{\sqrt6}}\,\left(q_i^{\rm T} C\gamma^5q_j\right)q_k^{\,\gamma}.
\end{equation}
The superscript T denotes a spinor transpose. We will use this compressed notation in later sections of the paper. 

We emphasize that the transformation properties of the quark operator above determine those of the most general effective octet field $B_{ijk}^{\,\gamma}(x)$. As noted later in Sec.\ \ref{subsec:dynamics}, $B_{ijk}^{\,\gamma}$ can be regarded more abstractly as the octet component $(C\gamma^5)_{\alpha \beta}\,\psi_{ijk}^{\,\alpha\beta\gamma}(x)$ of a general six-index interpolating field $\psi_{ijk}^{\,\alpha\beta\gamma}(x)$ which can be used to calculate Green's functions in the low-momentum limit of QCD. Moreover, the flavor and spin correlations that appear in matrix elements calculated using the quark operators determine the most general structure of matrix elements expressed in terms of the effective fields $\psi$ or $B$, a fact we will use extensively in later sections of the paper.

A representation similar to that in Eq.\ (\ref{Bijk}) was used by Labrenz and Sharpe \cite{Labrenz} in their study of quenched chiral perturbation theory for baryons. However, because of the presence of bosonic as well as fermionic quarks in their formalism for suppressing quark loops, they found it necessary to symmetrize in the last two flavor indices in Eq.\ (\ref{Bijk}) to eliminate a flavor-singlet component of their fields. There is no flavor singlet here, and our representation is correspondingly simpler. 

It is straightforward to show that 
\begin{equation}
\label{Bsymmetry}
B_{jik}^{\,\gamma} = -B_{ijk}^{\,\gamma},
\end{equation}
so the quarks $q_i$ and $q_j$ must have different flavors. There is no symmetry constraint with respect to those quarks and $q_k$. The absence of any flavor- and color-singlet combination of three quarks with total spin $j=1/2$ means that $\epsilon_{ijk}\,B_{ijk}^{\,\gamma} \equiv 0$ giving the Jacobi-type identity %
\begin{equation}
\label{Jacobi}
B_{ijk}^{\,\gamma} + B_{jki}^{\,\gamma} + B_{kij}^{\,\gamma} \equiv 0.
\end{equation}
These relations will be used extensively in later sections of the paper.

We have normalized $B_{ijk}^{\,\gamma}$ to correspond as follows to the octet baryons with the standard choice of baryon phases:
\renewcommand{\arraystretch}{1.3}
\begin{equation}
\label{octet}
\begin{array}{c}
B_{121} \leftrightarrow {\textstyle\frac{1}{\!\!\sqrt{2}}}\,p, \quad B_{122} \leftrightarrow {\textstyle\frac{1}{\!\!\sqrt{2}}}\,n ,\\ B_{131} \leftrightarrow {\textstyle\frac{1}{\!\!\sqrt{2}}}\,\Sigma^+, \quad B_{232} \leftrightarrow {\textstyle\frac{1}{\!\!\sqrt{2}}}\,\Sigma^-, \\ B_{231} \leftrightarrow {\textstyle\frac{1}{\!\!\sqrt{2}}}\,\Sigma^0 + {\textstyle\frac{1}{\!\!\sqrt{6}}}\,\Lambda, \quad B_{132} \leftrightarrow {\textstyle\frac{1}{\!\!\sqrt{2}}}\,\Sigma_0 -
{\textstyle\frac{1}{\!\!\sqrt{6}}}\,\Lambda, \\
B_{133} \leftrightarrow {\textstyle\frac{1}{\!\!\sqrt{2}}}\,\Xi^0, \quad B_{233} \leftrightarrow {\textstyle\frac{1}{\!\!\sqrt{2}}}\,\Xi^-\,. 
\end{array}
\end{equation}
\renewcommand{\arraystretch}{1}
The remaining $B$\,s can be identified with the baryons using the relations in Eqs.\ (\ref{Bsymmetry}) and (\ref{Jacobi}). With this normalization, we can sum over repeated indices in subsequent equations. Finally, the three-flavor-index tensors $B_{ijk}$ are related to the two-index matrix $B_{kl}$ defined in Eq.\ (\ref{baryons}) by
\begin{equation}
\label{Bkl}
B_{kl}^{\,\gamma} = {\textstyle\frac{1}{\!\!\sqrt{2}}}\, \epsilon_{ijl}\,B_{ijk}^{\,\gamma}, \quad 
B_{ijk}^{\,\gamma} = {\textstyle\frac{1}{\!\!\sqrt{2}}}\,\epsilon_{ijl}\,B_{kl}^{\,\gamma}.
\end{equation}

The interpretation of the fields $B_{ijk}^{\,\gamma}$ can be clarified by going to the Lorentz frame in which $v^\mu=(1,\,\bf{0})$, that is, the rest frame of the baryon with momentum $p^\mu=m_0v^\mu$. Up to corrections of order $k/m_0$, the matrix elements of the Dirac matrices between $B$'s reduce in this frame to matrix elements of the $2\times 2$ Pauli matrices, 
\renewcommand{\arraystretch}{1.5}
\begin{equation}
\label{eq:Dirac_reduction}
\begin{array}{c}
1 \longrightarrow \openone, \quad \gamma^5 \longrightarrow 0, \quad \gamma^\mu \longrightarrow v^\mu  \nonumber \\ \gamma^\mu\gamma^5  \longrightarrow (0,\, \mbox{\boldmath$\sigma$}), \quad \sigma^{0 \mu}  \longrightarrow 0, \quad \sigma^{ij} \longrightarrow \epsilon_{ijk}\sigma_k.
\end{array}
\end{equation}
\renewcommand{\arraystretch}{1}
In particular, using $C\gamma^5 \longrightarrow -i\sigma_2$, we find that 
\begin{equation}
\label{NRbaryon}
B_{ijk}^{\,\gamma} \longrightarrow -{\textstyle\frac{1}{6}}\epsilon_{abc} \left(q^{\rm T\,a}_ii\sigma_2q_j^{\,b}\right)q_k^{\,c,\gamma} \equiv-{\textstyle\frac{1}{\sqrt{6}}}\left(q^{\rm T}_ii\sigma_2 q_j\right)q_k^{\,\gamma} ,
\end{equation}
where $q_k^{\,\gamma}$ is now a two-component spinor with spin index $\gamma \in \textstyle\pm\frac{1}{2}$. The factor in parentheses is a standard representation of a singlet spin configuration of the spinors $q_i,\, q_j$. 

We can put $B_{ijk}^{\,\gamma}$ in a form that displays the structure of the SU(6) wave functions of the quark model by using the expansion of an arbitrary $2\times 2$ matrix over the complete set of Pauli matrices,
\begin{equation}
\label{Fierz}
A_{\gamma\,\beta} = {\textstyle\frac{1}{2}}\left(\delta_{\gamma\,\beta}\,{\rm Tr}\,A + (\mbox{\boldmath$\sigma$})_{\gamma\,\beta}\cdot {\rm Tr}\, \mbox{\boldmath$\sigma$}A\right)
\end{equation}
to rearrange the spinors to combine $q_i$ and $q_k$ in a spinor product. Choosing $A_{\gamma\,\beta}$ as 
\begin{equation}
\label{Agamma_beta}
A_{\gamma\,\beta}=q_k^{\,\gamma}\left(q_i^{\rm T}i\sigma_2\right)^\beta,
\end{equation}
 we obtain
\begin{equation}
B_{ijk}^{\,\gamma} = -{\textstyle\frac{1}{2\sqrt{6}}} \left[\left(q_i^{\rm T} i\sigma_2q_k\right)q_j^{\,\gamma} + \left(q_i^{\rm T}i\sigma_2 \mbox{\boldmath$\sigma$}q_k\right) \cdot \left(\mbox{\boldmath$\sigma$} q_j\right)^{\,\gamma}\right]\,,\quad i\not=j.
\end{equation}
The first term in this expression is antisymmetric in the indices $i,\,k$ and has those quarks in a singlet spin state. This term vanishes except for the $\Lambda$ hyperon, where $i,\,k\in u,\,d$. The second term has quarks $i$ and $k$ in a triplet spin state, is symmetric in those indices, and reproduces the expected SU(6) structure of the remaining octet baryons when written out in detail for a specific choice of $\gamma$. 

It is easily checked that the spin operator $\mbox{\boldmath$\sigma$}_i \!\cdot\mbox{\boldmath$ \sigma$}_j$ has the expected values -3 (1) when acting on a singlet (triplet) configuration of two quarks, where  $\mbox{\boldmath$\sigma$}_i$ denotes the action of the Pauli matrix $\mbox{\boldmath$\sigma$}$ on $q_i$. Thus,
\begin{eqnarray}
\label{sigma_sigma}
\mbox{\boldmath$\sigma$}_i\!\cdot\mbox{\boldmath$ \sigma$}_j\,q_i^{\rm T} \left(i\sigma_2\right)q_j &\equiv& \left(\mbox{\boldmath$\sigma$}q_i\right)^{\rm T}  \cdot\left(i\sigma_2 \mbox{\boldmath$\sigma$}q_j\right) = -3q_i^{\rm T}  \left(i\sigma_2\right)q_j, \\ \mbox{\boldmath$\sigma$}_i\!\cdot\mbox{\boldmath$ \sigma$}_j\,q_i^{\rm T} \left(i\sigma_2\right)\mbox{\boldmath$\sigma$}q_j &\equiv& (\sigma_lq_i)^{\rm T} \left(i\sigma_2\right)\mbox{\boldmath$\sigma$}\sigma_l q_j = q_i^{\rm T} \left(i\sigma_2\right)\mbox{\boldmath$\sigma$}q_j\,,
\end{eqnarray}
where we have used the relations
\begin{equation}
\label{transpose}
\mbox{\boldmath$\sigma$}^{\rm T}\left(i\sigma_2\right) = -\left(i\sigma_2\right)\mbox{\boldmath$\sigma$},
\end{equation}
$\mbox{\boldmath$\sigma$}\!\cdot \mbox{\boldmath$\sigma$}=3$, and $\sigma_l\mbox{\boldmath$\sigma$}\sigma_l=-\mbox{\boldmath$\sigma$}$ with an implied sum over the repeated index $l$ in the last.

We will also need, more generally, the action of $\mbox{\boldmath$\sigma$}_i \!\cdot\mbox{\boldmath$ \sigma$}_k$ on a mixed-symmetry combination $q_k^{\,\gamma}q_i^{\,\alpha}$. Using the expansion in Eq.\ (\ref{Fierz}), we can rewrite this product as
\begin{equation}
\label{qk-qi}
q_k^{\,\gamma}q_i^{\,\alpha} = {\textstyle\frac{1}{2}}\,[\delta_{\gamma\alpha} \,(q_i^{\rm T}q_k) + (\sigma_1)_{\gamma\alpha}\,(q_i^{\rm T}\sigma_1q_k) + (\sigma_2)_{\gamma\alpha}\,(q_i^{\rm T}\sigma_2q_k) + (\sigma_3)_{\gamma\alpha}\,(q_i^{\rm T}\sigma_3q_k)\,].
\end{equation}
The third term on the right hand side has the quarks in a singlet configuration. The remaining symmetrical combinations are triplets. As a consequence,
\begin{equation}
\label{sigma-sigma}
\mbox{\boldmath$\sigma$}_i \!\cdot\mbox{\boldmath$ \sigma$}_k\, q_k^{\,\gamma}q_i^{\,\alpha} = q_k^{\,\gamma}q_i^{\,\alpha} - 2(\sigma_2)_{\gamma\alpha}\,(q_i^{\rm T}\sigma_2q_k)\,,
\end{equation}
where we have added and subtracted a singlet term to reproduce the original product as the first term on the right. We find immediately that
\begin{equation}
\label{Pik-singlet}
(1-\mbox{\boldmath$\sigma$}_i \!\cdot\mbox{\boldmath$ \sigma$}_k)\,q_k^{\,\gamma}q_i^{\,\alpha} = 2\,(\sigma_2)_{\gamma\alpha}\,(q_i^{\rm T}\,\sigma_2\,q_k) = 2\,(i\sigma_2)_{\gamma\alpha}\,(q_k^{\rm T}\,i\sigma_2\,q_i),
\end{equation}
a relation we will need later. 

Alternatively, adding $q_k^{\,\gamma}q_i^{\,\alpha}$ to both sides of the expression in Eq.\ (\ref{sigma-sigma}) and evaluating the right hand side explicitly, we find that the operator
\begin{equation}
\label{Pij}
P_{ik} = {\textstyle\frac{1}{2}\left(1+\mbox{\boldmath$\sigma$}_i\!\cdot\mbox{\boldmath$ \sigma$}_k\right)}
\end{equation}
exchanges the spin indices of quarks $i$ and $k$ or, alternatively,
acts as the exchange operator for the flavor indices $i,\,k$ when the order of the spin indices is kept fixed, 
\begin{equation}
\label{exchange}
P_{ik}\,q_i^{\,\alpha} q_k^{\,\gamma} = q_i^{\,\gamma} q_k^{\,\alpha} = q_k^{\,\alpha} q_i^{\,\gamma}\,.
\end{equation}

We note finally that the projection operator for a total spin-1/2 configuration of three quarks is
\begin{equation}
\label{project_1/2}
P^{1/2} = {\textstyle\frac{1}{6}}\left(3 - \mbox{\boldmath$\sigma$}_i \!\cdot \mbox{\boldmath$ \sigma$}_j -\mbox{\boldmath$\sigma$}_j \!\cdot \mbox{\boldmath$ \sigma$}_k -\mbox{\boldmath$\sigma$}_k \!\cdot \mbox{\boldmath$ \sigma$}_i \right).
\end{equation}

\subsubsection{The decuplet-baryon fields}
\label{subsubsec:decuplet_3index}

The decuplet baryons are represented in the SU(6) language \cite{Pais} by a field $T^{\mu\gamma}_{ijk}$ which is a Rarita-Schwinger spinor with four-vector and spinor indices $\mu$ and $\gamma$, is a completely symmetric tensor in the flavor indices $i,j,k\in 1,2,3\equiv u,d,s$, and is a color singlet. In particular, $T$ transforms as the three-quark combination
\begin{equation}
\label{Tijk}
T^{\,\mu\gamma}_{ijk}={\textstyle\frac{1}{18\!\!\sqrt{2}}}\epsilon_{abc}\left(q_i^{\,\alpha a} q_j^{\,\beta b} q_k^{\,\gamma c} + q_k^{\,\alpha a} q_j^{\,\beta b} q_i^{\,\gamma c} + q_i^{\,\alpha a} q_k^{\,\beta b} q_j^{\,\gamma c}\right) \left(C\gamma^\mu\right)_{\alpha\beta},
\end{equation}
or equivalently, as the decuplet component 
\begin{equation}
\sum_{P(ijk)}\,\left(C\gamma^\mu\right)_{\alpha\beta}\psi_{ijk}^{\,\alpha\beta \gamma}
\end{equation}
of the general six-index interpolating field discussed in Sec.\ \ref{subsec:dynamics}. In particular, the transformation of $T_{ijk}$ under SU(3)$_V$ follows from Eqs.\ (\ref{eq:Qtrans}) and (\ref{Tijk}),
\begin{equation}
\label{delta_transform}
T_{ijk}^{\,\mu\gamma} \stackrel{U}{\longrightarrow} U_{ii'}U_{jj'}U_{kk'} T_{i'j'k'}^{\,\mu\gamma}.
\end{equation}

The Rarita-Schwinger constraint $\gamma_\mu T^{\mu\gamma}=0$ reduces to $v_\mu T^{\mu\gamma}=0$ in the rest frame of the baryon. As a result, $T^{\mu\gamma}\rightarrow\left(0,\,{\bf T}^\gamma \right)$ in that frame, where the spatial vector ${\bf T}^\gamma$ is given by
\begin{equation}
\label{T_restframe}
{\bf T}^\gamma ={\textstyle \frac{1}{18\sqrt{2}}} \epsilon_{abc}\left[\left(q_i^{{\rm T} \,a} i\sigma_2\mbox{\boldmath$\sigma$}q_j^b\right)q_k^{\gamma c} + j\leftrightarrow k + k\leftrightarrow i\right].
\end{equation}
Alternatively, with color suppressed,
\begin{equation}
\label{Trest_nocolor}
{\bf T}^\gamma ={\textstyle \frac{1}{6\sqrt{3}}} \left[\left(q_i^{\rm T} i\sigma_2\mbox{\boldmath$\sigma$}q_j\right)q_k^{\,\gamma} + j\leftrightarrow k + k\leftrightarrow i\right]\,,
\end{equation}
an expression which can be written as
\begin{equation}
\label{Tproj}
{\bf T}^\gamma = P^{3/2}\,{\textstyle\frac{1}{\!2\sqrt{3}}}\,\left(q_i^{\rm T} i\sigma_2\mbox{\boldmath$\sigma$}q_j\right)q_k^{\,\gamma}\,.
\end{equation}
Here $P^{3/2}$ is the projection operator for total spin 3/2,
\begin{eqnarray}
\label{project_3/2}
P^{3/2} &=& {\textstyle\frac{1}{6}}\left(3 + \mbox{\boldmath$\sigma$}_i\! \cdot \mbox{\boldmath$\sigma$}_j + \mbox{\boldmath$\sigma$}_j \!\cdot \mbox{\boldmath$\sigma$}_k + \mbox{\boldmath$\sigma$}_k \!\cdot \mbox{\boldmath$\sigma$}_i\right) \nonumber \\ &=&{\textstyle\frac{1}{3}} (\, P_{ij} + P_{jk} + P_{ki}\,),
\end{eqnarray}
with $P_{ij}$ the flavor permutation operator defined in Eq.\ (\ref{Pij}).

The field ${\bf T}_{111}^\gamma = \left(T_{111}^{1,\gamma},\, T_{111}^{2,\gamma},\, T_{111}^{3,\gamma}\right)$ is normalized to the $\Delta^{++}$, with the combination
\begin{eqnarray}
\label{Delta++}
\Delta^{++}_{j_z=3/2} &=& {\textstyle\frac{1}{\!\!\sqrt{2}}} \left(T^{1,1/2}_{111}-iT^{2,1/2}_{111}\right)\\
&=& {\textstyle\frac{1}{3\sqrt{6}}}\left[\left(q_i^{\rm T}i\sigma_2\sigma^-q_j\right)q_k^{\,1/2} + j\leftrightarrow k + k\leftrightarrow i\right]
\end{eqnarray}
acting as the annihilation operator for the $\Delta^{++}$ state with $j_z=+3/2$. Here $\sigma^\pm$ are the usual spin raising and lowering operators,
\begin{equation}
\label{sigma_pm}
\sigma^\pm = {\textstyle\frac{1}{2}}(\sigma_1\pm i\sigma_2).
\end{equation}

The remaining decuplet baryons with $j_z=+3/2$ have the same spin structure. The replacement of $T_{111}$ by $T_{ijk}$ in Eq.\ (\ref{Delta++}) gives the following connections:
\renewcommand{\arraystretch}{1.3}
\begin{equation}
\label{decuplet}
\begin{array}{c}
T_{111} \leftrightarrow \Delta^{++},\quad T_{112} \leftrightarrow {\textstyle\frac{1}{\!\!\sqrt{3}}}\,\Delta^+, \quad T_{122} \leftrightarrow {\textstyle\frac{1}{\!\!\sqrt{3}}}\,\Delta^0,\quad T_{222} \leftrightarrow \Delta^-\\ T_{113} \leftrightarrow {\textstyle\frac{1}{\!\!\sqrt{3}}}\,\Sigma^{*+}, \quad T_{123} \leftrightarrow {\textstyle\frac{1}{\!\!\sqrt{6}}}\,\Sigma^{*0},\quad T_{223} \leftrightarrow {\textstyle\frac{1}{\!\!\sqrt{3}}}\,\Sigma^{*-} \\ T_{133} \leftrightarrow {\textstyle\frac{1}{\!\!\sqrt{3}}}\,\Xi^{*0},\quad T_{233} \leftrightarrow {\textstyle\frac{1}{\!\!\sqrt{3}}}\,\Xi^{*-}, \quad T_{333} \leftrightarrow \Omega^-.
\end{array}
\end{equation}
\renewcommand{\arraystretch}{1}
The remaining $T$\,s are determined by the complete symmetry in the flavor indices.

\subsubsection{Pseudoscalar meson fields}
\label{subsubsec:meson_representation}

The effective octet pseudoscalar meson fields $\phi_{ij}$ correspond to quark-antiquark pairs in a singlet spin configuration,
\begin{equation}
\phi_{ij} = {\textstyle\frac{1}{\!\!\sqrt{6}}}\,\left(q_i^{\,\alpha a}\bar{q}_j^{\, \beta b} - {\textstyle\frac{1}{3}}\delta_{ij}\, q_k^{\,\alpha a}\bar{q}_k^{\, \beta b}\right)\,\delta_{ab}\,(C\gamma^5)_{\alpha \beta}
\end{equation}
and transform under SU(3)$_V$ as
\begin{equation}
\label{meson_trans}
\phi_{ij} \stackrel{U}{\longrightarrow} U_{ii'}U^*_{jj'}\phi_{i'j'} = U_{ii'}\phi_{i'j'}U^\dagger_{j'j}.
\end{equation}
The use of this representation makes the quark flow in an interaction diagram clear. However, physical meson masses are customarily used in loop calculations in chiral perturbation theory so we will generally use the representation
\begin{equation}
\phi_{ij} = {\textstyle\frac{1}{2}}\,\sum_{l=1}^8\,\lambda_{ij}^l\phi^{\,l}
\end{equation}
instead, with
\begin{equation}
\phi^{\,l} = \sum_{ij}\,\lambda_{ji}^l\phi_{ij} = {\textstyle\frac{1}{\!\!\sqrt{6}}}\,\sum_{ij}\,\lambda_{ij}^lq_i^{\,\alpha a}\bar{q}_j^{\, \beta b}\,\delta_{ab}\,(C\gamma^5)_{\alpha \beta}.
\end{equation}
Note that we will not include the flavor-singlet pseudoscalar, nominally the $\eta'$, in our later calculations.

\subsection{Baryon and meson propagators}
\label{subsec:propagators}

The momentum-space propagators for $B$, $T$, and $\phi$ can be constructed fairly easily. There is an overall factor 
\begin{equation}
\label{baryon_prop}
i/(v\cdot k + i\epsilon),
\end{equation}
for a heavy baryon with momentum $p^\mu = m_0v^\mu + k^\mu$ \cite{georgi-HBPT,Jen-HBChPT}, and the usual factor 
\begin{equation}
i/(k^2-M^2+i\epsilon)
\end{equation}
for a meson with momentum $k^\mu$ and mass $M$. The spin and flavor structures are easily determined in the quark representation by calculating vacuum expectation values of products of the fields and their conjugates expressed in terms of the $q$'s and $\bar{q}$'s. Thus, in a heavy baryon, 
\begin{eqnarray}
\label{quark_prop}
\left<0|\,q_{i'}^{\,\alpha a}\bar{q}_i^{\,\beta b}\,|0\right> &=& {\textstyle \frac{1}{2}}\,(1+\not\! v)^{\alpha\beta}\,\delta^{ab}\,\delta_{i'i} \\ 
&\mbox{$\longrightarrow$}& \: \delta^{\alpha\beta}\,\delta^{ab}\,\delta_{i'i}, \quad v_\mu \longrightarrow (1, {\bf 0})\, ,
\end{eqnarray}
where the quark must be taken to move with the four velocity $v^\mu$ of the baryon with
\begin{equation}
(\not\! v - 1)\,q=0.
\end{equation} 

The results for the propagators from an intial state with flavors $i,\,j,\,k$ and spinor and Lorentz indices $\gamma,\,\mu$ to a final state with flavors $i',\,j',\,k'$ and indices $\gamma',\,\mu'$ are
\begin{eqnarray}
\label{Bpropagator}
B:\quad && iS^{\,\gamma'\gamma}_{B;i'j'k';kji}(v,k) = P^{\,B\,\gamma'\gamma}_{i'j'k';kji}\,\frac{i}{v.k+i\epsilon}\,, \\
\label{Tpropagator}
T:\quad && iS^{\,\mu'\gamma';\mu\gamma}_{T;i'j'k';kji}(v,k) = P^{\,T\,\mu'\gamma';\mu\gamma}_{i'j'k';kji}\,\frac{i}{v.k+i\epsilon} \,, \\
\label{phi_propagator}
\phi:\quad && i\Delta_{j'j}(k) = \delta_{j'j}\,\frac{i}{k^2-M^2+i\epsilon}\,.
\end{eqnarray}
The projection operators for the octet and decuplet baryons are
\begin{eqnarray}
\label{projection_op_oct}
P^{B;\gamma'\gamma}_{i'j'k';kji} &=&  P_{1/2}^{\,\gamma'\gamma}\, 
\label{projection_op_dec}
P^B_{i'j'k';kji}\,, \\
P^{T;\mu'\gamma';\mu\gamma}_{i'j'k';kji} &=& P^{\,\mu'\gamma';\mu\gamma}_{3/2}\, P^T_{i'j'k';kji}\,,
\end{eqnarray}
where $P_{1/2}$ and $P_{3/2}$ are the spin projection operators for spins 1/2 and 3/2,
\begin{eqnarray}
\label{1/2proj}
P_{1/2}^{\,\gamma'\gamma} &=& {\textstyle\frac{1}{2}} (1+\not\! v)_{\gamma'\gamma}\,, \\
\label{3/2proj}
P_{3/2}^{\,\mu'\gamma';\mu\gamma} &=& \left[\,{\textstyle\frac{1}{2}}(1+\not\! v)\,   
\left(v^{\,\mu'}\!v^{\,\mu} - g^{\,\mu'\mu} - {\textstyle\frac{4}{3}}S^{\mu'}\!S^\mu\right)\,\right]_{\gamma'\gamma}\,
\end{eqnarray}
and $P^B$ and $P^T$ are flavor projection operators
\begin{eqnarray}
\label{PB}
P^B_{i'j'k';kji} &=& {\textstyle\frac{1}{6}}\left(2\delta_{i'i}\delta_{j'j} \delta_{k'k} - 2 \delta_{i'j}\delta_{j'i} \delta_{k'k} + \delta_{i'i}\delta_{j'k} \delta_{k'j} \right. \nonumber \\
&&\left. -\; \delta_{i'j}\delta_{j'k} \delta_{k'i} + \delta_{i'k}\delta_{j'j} \delta_{k'i}- \delta_{i'k}\delta_{j'i} \delta_{k'j} \right)\,,\\
\label{PT}
P^T_{i'j'k';kji} &=& {\textstyle\frac{1}{6}}\left(\delta_{i'i}\delta_{j'j} \delta_{k'k} + \delta_{i'j}\delta_{j'i} \delta_{k'k} + \delta_{i'i}\delta_{j'k} \delta_{k'j}\right. \nonumber\\
&&\left. +\; \delta_{i'j}\delta_{j'k} \delta_{k'i} + \delta_{i'k}\delta_{j'j} \delta_{k'i}+ \delta_{i'k}\delta_{j'i} \delta_{k'j} \right)\,.
\end{eqnarray}

The projection operators have the properties 
\begin{eqnarray}
\label{Bprojection}
P^{\,B\,\gamma'\gamma}_{i'j'k';kji}\,B_{ijk}^{\,\gamma'} &=& B_{i'j'k'}^{\,\gamma'}, \qquad \bar{B}_{k'j'i'}^{\,\gamma'}\,P^{\,B\,\gamma'\gamma}_{i'j'k';kji} = \bar{B}_{kji}^{\,\gamma}, \\
\label{Tprojection}
P^{\,T\,\mu'\gamma';\mu\gamma}_{i'j'k';kji}\,T_{ijk}^{\,\mu\gamma} &=& T_{i'j'k'}^{\,\mu'\gamma'}, \qquad \bar{T}_{k'j'i'}^{\,\mu\gamma}\,P^{\,T\,\mu'\gamma';\mu\gamma}_{i'j'k';kji} = \bar{T}_{kji}^{\,\mu\gamma},
\end{eqnarray}
where repeated indices are to be summed, both here and later. Our labelling conventions are such that
\begin{eqnarray}
\label{BBbar}
P^{\,B\,\gamma'\gamma}_{i'j'k';kji} &=& \left<k'j'i';\gamma'|ijk;\gamma \right>,\\ P^{\,T\,\mu'\gamma';\mu\gamma}_{i'j'k';kji} &=&  \left<k'j'i';\mu'\gamma'|ijk;\mu\gamma \right>.
\end{eqnarray}

The results for the propagators and projection operators are easy to derive using either the covariant representation of the heavy-baryon fields in Eqs.\ (\ref{Bijk}) and (\ref{Tijk}), or the rest-frame representations in Eqs.\ (\ref{NRbaryon}) and (\ref{Trest_nocolor}) and time-ordered perturbation theory \cite{Weinberg}. Because the production of baryon-antibaryon pairs ($Z$ graphs) vanishes in the heavy-baryon limit, the propagator factor $1/v\cdot k$ is equivalent to the  simple energy denominator $1/[E(m_0v+k)-E(m_0v)]$. The spin projection operators are given directly by sums over intermediate spin states in the baryon rest frame,  with 
\begin{eqnarray}
\label{NRspin_proj}
P_{1/2}^{\,\gamma'\gamma} &=& \delta_{\,\gamma'\gamma}, \\
P_{3/2}^{\,r'\gamma';r\gamma} &=& {\textstyle\frac{1}{3}}\left[ \delta_{\,r'r}\delta_{\,\gamma'\gamma} + (\sigma_r\sigma_{r'})_{\gamma'\gamma} \right] \nonumber \\
&=& \delta_{\,r'r}\delta_{\gamma'\gamma} - {\textstyle\frac{1}{3}} (\sigma_{r'}\sigma_{r})_{\gamma'\gamma} \\
&=& \delta_{\,r'r}\delta_{\,\gamma'\gamma} - {\textstyle\frac{4}{3}} (S_{r'}S_{r})_{\gamma'\gamma} \,, \nonumber
\end{eqnarray}
in that frame. The indices $r,\,r'$ in $P_{3/2}^{\,r'\gamma';r\gamma}$ label the components $T^{r\gamma}$ of the spatial vector ${\bf T}^\gamma$, Eq.\ (\ref{T_restframe}). The projection operators in Eqs.\ (\ref{1/2proj}) and (\ref{3/2proj}) are the covariant generalizations of these results. 

We will also need the projection operator that extracts the spin-1/2 component of a vector-spinor product. This is given by
\begin{equation}
P^{\,r'\gamma';r\gamma}_{1/2}= {\textstyle\frac{1}{3}}(\sigma_{r'} \sigma_r)_{\gamma'\gamma} = {\textstyle\frac{4}{3}}(S_{r'}S_r)_{\gamma'\gamma},
\end{equation}
with
\begin{equation}
P^{\,r'\gamma';r\gamma}_{1/2} + P_{3/2}^{\,r'\gamma';r\gamma} = \delta_{r'r}\delta_{\gamma'\gamma}.
\end{equation}
These operators hold for an general vector-spinor product. They are equivalent to the operators $P^{1/2}$ and $P^{3/2}$ in Eqs.\ (\ref{project_1/2}) and (\ref{project_3/2}) for products expressed in terms of quark fields.\footnote{For example,
\begin{eqnarray*}
P_{3/2}^{\,r'\gamma';r\gamma}\left(q_i^{\,\rm T}i\sigma_2\sigma_r q_j\right) q_k^{\,\gamma} &=& {\textstyle\frac{1}{3}}\left(q_i^{\,\rm T}i\sigma_2\sigma_{r'} q_j\right)q_k^{\,\gamma'} + {\textstyle\frac{1}{6}} \left(q_i^{\,\rm T}i\sigma_2\sigma_r q_j + q_j^{\,\rm T}i\sigma_2\sigma_r q_i\right)\left(\sigma_r\sigma_{r'}q_k\right)^{\gamma'} \\
&=& {\textstyle\frac{1}{3}}\left[\left(q_i^{\,\rm T}i\sigma_2\sigma_{r'} q_j\right)q_k^{\,\gamma'} + \left(q_i^{\,\rm T}i\sigma_2\sigma_{r'} q_k\right)q_j^{\,\gamma'} + \left(q_j^{\,\rm T}i\sigma_2\sigma_{r'} q_k\right)q_i^{\,\gamma'} \right]\\ &=& {\textstyle\frac{1}{3}}\left(P_{ij} + P_{jk} + P_{ki}\right)\left(q_i^{\,\rm T}i\sigma_2\sigma_{r'} q_j\right) q_k^{\,\gamma'} = P^{3/2}\left(q_i^{\,\rm T}i\sigma_2\sigma_{r'} q_j\right) q_k^{\,\gamma'}\,,
\end{eqnarray*}
where we have used the symmetry of $q_i^{\,\rm T}i\sigma_2\sigma_r q_j$ in the first line, and have then used the relations in Eqs.\ (\ref{sigma-sigma}) and (\ref{project_3/2}) and the antisymmetry of  $q_i^{\,\rm T}i\sigma_2 q_j$ to reduce the result.
}

\section{CHIRAL INTERACTIONS IN THREE-FLAVOR-INDEX FORM}
\label{sec:chiral-interactions}

\subsection{The chiral effective Lagrangian}
\label{subsec:interactions}


It is straightforward to rewrite the chiral Lagrangian in Eq.\ (\ref{lag}) in the three-flavor-index notation. The transformations of the fields $B_{ijk}^{\,\gamma}$ and $T_{ijk}^{\,\mu\gamma}$ under general $\gamma^5$ transformations implied by Eqs.\ (\ref{baryon_transform}) and (\ref{delta_transform}) lead to the covariant derivatives
\begin{eqnarray}
\label{covar_derivative}
{\cal D_\mu}B_{ijk}^{\,\gamma} &=& \partial_\mu B_{ijk}^{\,\gamma} + V_{ii'}B_{i'jk}^{\,\gamma} + V_{jj'}B_{ij'k}^{\,\gamma} + V_{kk'}B_{ijk'}^{\,\gamma},\\{\cal D_\mu}T_{ijk}^{\,\nu\gamma} &=& \partial_\mu T_{ijk}^{\,\nu\gamma} + V_{ii'}T_{i'jk}^{\,\nu\gamma} + V_{jj'}T_{ij'k}^{\,\nu\gamma} + V_{kk'}T_{ijk'}^{\,\nu\gamma},
\end{eqnarray}
where the $V$'s are the components of the vector current matrix in Eq.\ (\ref{Vcurrent}). The leading-order Lagrangian becomes
\begin{eqnarray}
\label{L0_B,T}
 {\cal L}_0 &=&  i\,(\bar{B} v\!\cdot\!{\cal D}\, B) - \delta m_B\, (\bar{B}B)
+ 2(D+F)\, (\bar{B} S^\mu \! B A_\mu)
- 4(D-F)\, (\bar{B} S^\mu\! A_\mu B) \nonumber \\
&& -i\,(\bar{T}^\mu v\!\cdot\!{\cal D} T_\mu)
+ \delta m_T\,(\bar{T}^\mu T_\mu) 
+ 2\,{\cal H}\, (\bar{T}^\mu S^\nu \! A_\nu T_\mu)\\
&& +\sqrt{2}\,{\cal C} \left[ (\bar{T}^\mu A_\mu B) + (\bar{B} A_\mu T^\mu) \right] + {\textstyle\frac{1}{4}}f^2\,(\partial_\mu \Sigma\,\partial^\mu\Sigma) \,, \nonumber
\end{eqnarray}
where the bilinear invariants in this representation are defined in general as 
\begin{mathletters}
\begin{eqnarray}
  (\bar{B}\,\Gamma{B}) &\equiv& 
\bar{B}_{kji}^{\,\alpha}\Gamma_{\alpha\beta} 
		{B}_{ijk}^{\,\beta}\,,
\label{eq:inv1}\\
  (\bar{B}\,\Gamma{B} A) 
	&\equiv& \bar{B}_{k'ji}^{\,\alpha}\Gamma_{\alpha\beta}
                                A_{k'k} {B}_{ijk}^{\,\beta}\,,
\label{eq:inv2}\\
  (\bar{B}\,\Gamma AB) &\equiv& 
\bar{B}_{kji'}^{\,\alpha}\Gamma_{\alpha\beta}
				A_{i'i} {B}_{ijk}^{\,\beta}\,,
\label{eq:inv3}\\
  (\bar{T}^{\,\mu}\Gamma{T}_\mu) &\equiv&
  \bar{T}^{\mu\alpha}_{kji} g_{\mu\nu} \Gamma_{\alpha\beta} 
	T_{ijk}^{\,\nu\beta} \,,
\label{eq:Tinv1}\\
  (\bar{T}^{\,\mu}\Gamma A^\lambda T_\mu) &\equiv&
 \bar{T}^{\,\mu\alpha}_{kji'}g_{\mu\nu} \Gamma_{\alpha\beta} 
                            A_{i'i}^\lambda T_{ijk}^{\,\nu\beta} \,,
\label{eq:Tinv2}\\
  (\bar{B} \Gamma A^\mu T_\mu) &\equiv&
 \bar{B}_{kji'}^{\,\alpha} g_{\mu\nu} \Gamma_{\alpha\beta} 
                            A_{i'i}^\mu T_{ijk}^{\,\nu\beta} \,,
\label{eq:Tinv3}\\
(\Sigma \Sigma) &\equiv& \Sigma_{ji}\Sigma_{ij} \,.
\label{Eq:sigma_invar}
\end{eqnarray}
\end{mathletters}
Here $\Gamma$ is an arbitrary Dirac matrix, $A$ is a scalar or vector operator, and $g_{\mu\nu}$ is the Lorentz metric tensor with signature -2. 

The couplings in Eq.\ (\ref{L0_B,T}) are the same as in the matrix form of ${\cal L}_0$ in Eq.\ (\ref{lag}). However, this Lagrangian could have been written down directly, with arbitrary coefficients, as the most general allowed by chiral invariance at the leading order in the derivative expansion.  

The expressions for ${\cal L}_0$ in Eqs.\ (\ref{lag}) and (\ref{L0_B,T}) are connected through the relations \footnote{Labrenz and Sharpe \cite{Labrenz} use a similar notation, but include an extra term in their definition of the field $B_{ijk}$ which is unnecessary in the present context. As a result their relations analogous to Eqs.\ (\ref{BbarBA}-\ref{BbarAT})are more complicated. } 
\begin{mathletters}
\begin{eqnarray}
\label{BbarB}
(\bar{B}\Gamma B) &=& ({\rm Tr}\,\bar{B}B)^{\alpha\beta}\,\Gamma_{\alpha \beta}, \\
\label{BbarBA}
(\bar{B}\Gamma BA) &=& ({\rm Tr}\,\bar{B}AB)^{\alpha\beta}\,\Gamma_{\alpha \beta}, \\
\label{BbarAB}
(\bar{B}\Gamma AB) &=& -{\textstyle\frac{1}{2}}({\rm Tr}\,\bar{B} BA)^{\alpha\beta}\,\Gamma_{\alpha \beta} + {\textstyle\frac{1}{2}}({\rm Tr}\,A)\;({\rm Tr}\,\bar{B}B)^{\alpha\beta}\,\Gamma_{\alpha \beta}, \\
\label{BbarAT}
(\bar{B }\Gamma A^\mu T_\mu) &=& {\textstyle\frac{1}{\!\!\sqrt{2}}}\, \epsilon_{i'jl}\bar{B}_{lk}^{\,\alpha}A_{i'i}^\mu T_{ijk}^{\,\mu\beta}\, \Gamma_{\alpha\beta}\,,\\
\label{SS}
(\Sigma\Sigma) &=& {\rm Tr}\,\Sigma\Sigma\,.
\end{eqnarray}
\end{mathletters}
The first three relations follow from Eq.\ (\ref{Bkl}). The spinor indices $\alpha$ and $\beta$ refer to $\bar{B}$ and $B$, and the traces on the right hand sides of these expressions to matrix traces with $\phi$ and  $B$  represented as in Eqs.\ (\ref{pions}) and (\ref{baryons}). ${\rm Tr}\,A_\mu=0$ for the axial current so the second term on the right hand side of Eq.\ (\ref{BbarAB}) vanishes for the invariant $(\bar{B}S^\mu \! A_\mu \!B)$. 

While the Lagrangian above gives a chiral description of baryonic processes in the low-momentum limit, it is not clear how the various terms in ${\cal L}_0$ are connected to the underlying dynamical theory, and in particular, what dynamical relations, if any, there may be among the effective couplings. We will explore aspects of this connection in the following sections, and will show that the strong couplings $D$, $F$, $\cal C$, and $\cal H$ in fact have the familiar SU(6) ratios when the spin-spin interactions in the dynamical theory are weak.

\subsection{Connection with dynamical models}
\label{subsec:dynamics}

The quark picture used above appears at one level as just a calculational device for keeping track of the flavor and spin indices of the most general effective baryon and meson fields  $B^{\,\gamma}_{ijk}(x)$, $T^{\,\mu\gamma}_{ijk}(x)$, and $\phi_{ij}(x)$. However, at a deeper level, the hadrons are dynamical quark-gluon systems with currents and interactions defined at the quark level, including the symmetry-breaking quark mass terms
\begin{equation}
\label{quark_masses}
\bar{q}_i^{\,a}m_iq_i^{\,a}
\end{equation}
in the basic QCD Lagrangian. The matrix elements of quark-level operators in hadronic states involve averages over the internal structures of the hadrons, with only the spin and flavor indices of the external particles being left at the end to label the matrix elements. This structure of the matrix elements is simply parametrized in HBChPT through the effective interactions of point hadrons given in Eq.\ (\ref{L0_B,T}), with the couplings representing the unknown matrix elements. The internal quark-gluon structure of the hadrons appears only through the chiral momentum expansion. 

Dynamical models provide further information which reflects the underlying quark-gluon structure of the theory. To make this more explicit, we note that the gauge-invariant operators
\begin{equation}
\label{psi_interpolate}
\Psi_{ijk}^{\,\alpha\beta\gamma}(x_i,x_j,x_k,x) = N\,\epsilon_{a'b'c'}\,q_i^{\,\alpha a}(x_1)\,q_j^{\,\beta b}(x_2)\,q_k^{\,\gamma c}(x_3)\,U^{aa'}(x_1,x)U^{bb'}(x_2,x)U^{cc'}(x_3,x)
\end{equation}
are possible interpolating operators for color-singlet baryonic states with spinor indices $\alpha,\,\beta,\,\gamma$ containing dynamical quarks with flavors $i,\,j,\,k$. The factors $U$ are path-ordered integrals of the gauge potential (Wilson lines),  
\begin{equation}
\label{Uaa'}
U^{aa'}(x_1,x) = P\,\exp\left(i\int_x^{x_1}dx^{\mu} A_\mu (x)\right)\,.
\end{equation}
The $x$'s and the integration paths lie on a spacelike surface. These operators, or the operators $\psi_{ijk}^{\,\alpha\beta\gamma}(x)$ obtained by integrating the product of $\Psi_{ijk}^{\,\alpha\beta\gamma}(x_i,x_j,x_k,x)$ with an appropriate compact function over the $x_i$ on the surface, can be used to define Green's functions for the theory and to identify the physical states \cite{Nishijima}. The external hadronic states are still characterized by their spins and the flavors of the quarks. 

It is clear that a sufficiently low-momentum probe will not be sensitive to the detailed structure of the hadron. The small effect on matrix elements of non-zero $x_i$ in the interpolating field or the hadron wave function can be treated perturbatively in a expansion in the probe momentum around zero, that is, as a derivative expansion. A familiar example is given by the low-momentum expansion of the electromagnetic form factors of a composite system.

Brambilla, Consoli, and Prosperi \cite{brambilla} used a Green's function construction based on the operators $\Psi_{ijk}^{\,\alpha\beta\gamma}(x_i,x_j,x_k,x)$ to derive an effective semi-relativistic Hamiltonian for heavy quarks in quenched QCD. The structure of this Hamiltonian reflects the underlying dynamics, so it will be useful to write it down in part:
\begin{eqnarray}
\label{H_brambilla}
H &=& H_0 + \frac{\alpha_s}{3m_1^2} {\bf S}_1 \cdot \left[({\bf r}_{12}\times{\bf p}_1)\frac{1}{r_{12}^3} + ({\bf r}_{13}\times{\bf p}_1)\frac{1}{r_{13}^3} \right] \nonumber \\
&&-\frac{2\alpha_s}{3m_1m_2}\,\frac{1}{r_{12}^3}\, {\bf S}_1 \cdot ({\bf r}_{12}\times{\bf p}_2)  - \frac{2\alpha_s}{3m_1m_3}\,\frac{1}{r_{13}^3 }\, {\bf S}_1 \cdot ({\bf r}_{13}\times{\bf p}_3) + \cdots \\ 
&& + \frac{2\alpha_s}{3m_1m_2}\,\frac{1}{r_{12}^3}\left[3\,({\bf S}_1 \cdot \hat{r}_{12}) ({\bf S}_2 \cdot \hat{r}_{12}) - {\bf S}_1 \cdot {\bf S}_2 \right] + \frac{2\alpha_s}{3m_1m_2}\,\frac{8\pi}{3}\,\delta^3({\bf r}_{12}) {\bf S}_1 \cdot {\bf S}_2 + \cdots\,, \nonumber
\end{eqnarray}
where ${\bf r}_{ij} = {\bf x}_i - {\bf x}_j$, and ${\bf S}_i=\mbox{\boldmath $\sigma$}_i/2$. $H_0$ contains the kinetic terms and a spin-independent but velocity-dependent interaction $V_{SI}$,
\begin{equation}
\label{H0}
H_0 = \sum_{i=1}^3\sqrt{{\bf p}_i^2+m_i^2} + V_{SI}.
\end{equation}
In Eq.\ (\ref{H_brambilla}) we have displayed only the spin-orbit, tensor, and spin-spin interactions for particle 1 associated with the exchange of gluons between pairs of quarks. The ellipsis contain additional Thomas-type spin-orbit terms associated with the long-range part of the potential and the remaining spin-dependent terms obtained by cyclic permutations of the quark labels $1,\,2,\,3$. The full expression is given in \cite{brambilla}. The kinematic masses that appear are to be interpreted as the effective masses of dressed quarks while the factors $1/m_im_j$ are more properly non-local energy operators $1/E_iE_j$ that smear out the short-range singularities.

This Hamiltonian, obtained by other means and without the small velocity-dependent parts of $V_{SI}$, was used by Carlson, Kogut, and Pandaripande \cite{kogut} and Capstick and Isgur \cite{isgur} in successful fits of the observed spectra of the low-lying baryons with excitation energies up to about 1.4 GeV. We will interpret this success as showing that the basic structure of the interaction in Eq.\ (\ref{H_brambilla}) is correct as far as its spin dependence and the relative sizes of the various terms are concerned. In particular, the spin-dependent terms are generally small and can be treated as perturbations   

To see the structure that might be expected in the chiral expansion, it is useful to study $H$ in more detail starting in the limit in which there are no symmetry-breaking quark mass terms in the underlying Lagrangian. The effective masses of the quarks in Eq.\ (\ref{H_brambilla}) must then all be equal, and the Hamiltonian is completely symmetric in the three quarks. The unperturbed states defined by the spin-independent Hamiltonian $H_0$ are independent of the spin structure. $H_0$ is rotationally symmetric in the space variables, and the ground state of the system has total spatial angular momentum $L=0$ and a spatial wave function that is completely symmetric in the coordinates and is the same for all the octet and decuplet baryons. As a result, dynamical matrix elements that differ only in the baryons involved will be equal in the symmetrical limit up to the known effects of the spin wave functions.

Taken together, the octet and decuplet states defined with respect to $H_0$, hence also the fields $B_{ijk}$ and $T_{ijk}$, determine a \mbox{\boldmath$56$} representation of the spin-flavor SU(6) \cite{Jen-axial}.\footnote{The combination of the pseudoscalar mesons $\phi_{ij}$ and the vector meson $\rho^\mu_{ij}=\frac{1}{\sqrt{6}}\,\left(q_i^{\,\alpha a}\bar{q}_j^{\,\beta b}- \frac{1}{3}\delta_{ij}q_k^{\,\alpha a}\bar{q}_k^{\,\beta b}\right)\,\delta_{ab}\left(C\gamma^\mu\right)_{\alpha\beta}$ would give a \mbox{\boldmath$35$} of SU(6) except for the large mass difference associated with spin effects and the role of the pseudoscalars as would-be Goldstone bosons. The potential symmetry is further broken by quark mass differences.} The symmetry is broken perturbatively in matrix elements by the changes in the space and spin structure caused by the small spin-dependent interactions in Eq.\ (\ref{H_brambilla}) or other effects of short-distance gluon exchange. For example, the first-order change in the ground state energy associated with the spin-dependent terms is simply proportional to $\left<0|\sum_{i<j}{\bf S}_i\cdot{\bf S}_j|0\right>$, a structure that gives an octet-decuplet mass difference but does not remove the mass degeneracies within the multiplets.\footnote{When the hyperfine-type interaction is properly smeared out spatially as in \cite{isgur} rather than being treated as a delta-function interaction, there is a noticable shift in the octet-decuplet mass splitting in higher orders because of the different signs of the interaction for the two multiplets. The overall structure is not changed.} While the spin-dependent terms in the Hamiltonian generate a nontrivial spin structure in the wave functions, this only affects the baryon masses and moments at second order in those interactions \cite{phuoc-diss}. 

The situation changes if the quark masses are not equal. We will suppose that $m_u=m_d=0$ and $m_s\not=0$.  The O($m_s$) changes in the baryon masses are related through the Feynman-Hellman theorem \cite{feynman} to matrix elements of the corresponding O($m_s$) changes in the Hamiltonian. These are of two types. The changes associated with the kinetic terms are spin independent and involve only one quark at a time, that is, involve one-body operators. Those associated with the ``hyperfine'' term $\sum_{i<j}{\bf S}_i\cdot{\bf S}_j$ are two-body operators which involve spin couplings between two quarks. The changes in the spin-orbit couplings potentially involve all three quarks but again only appear at second order in the perturbation expansion in $\alpha_s$ and can be neglected. There are no general theorems on the O($m_s^2$) changes in the baryon masses. 

The results for the magnetic moments of the ground-state baryon are similar. However, there is no analog of the Feynman-Hellman theorem, and the observation that the moments change to first-order only through one-body operators depends on the approximate decoupling of the spatial and spin parts of the ground-state wave functions.

We will show in the following sections that this general structure carries over to the matrix elements parametrized in the chiral expansion for baryon masses and moments. This structure is essentially kinematic. In particular, the three-flavor-index description of the baryon fields through the $B$'s and $T$'s gives the most general labeling of the flavor and spin content of the external baryons. Since the quark or flavor lines are continuous, both dynamically and in the effective field theory, the initial flavor indices can be followed through a process to determine the final flavor indices, including any effects of the meson field $\phi$. 

The correlation of initial and final spins is more complicated because quark spins can be flipped by dynamical interactions, and there is no continuity requirement for the spin projection associated with a given flavor line. However, we will show that the spin structure of the transition operators in two-baryon transitions is completely described to O($m_s$) by the action of the identity operator, single spin operators \mbox{\boldmath$\sigma$}, the two-body spin-spin operators $\mbox{\boldmath$\sigma$}_m \!\cdot  \mbox{\boldmath$\sigma$}_n$, and a spin-independent mass operator ${\cal M}\propto m_s$.    

The addition of dynamical information allows us to sharpen our conclusions. We will assume, as discussed above, that spin-dependent interactions are relatively unimportant in determining the structure of the ground-state baryons, and that an expansion in powers $m_s$ is legitimate. In the absence of spin-exchange interactions, the spin structure of dynamical matrix elements would be determined completely by the spins of the external baryons and the structure of the elementary quark-level operators involved. At leading order in the derivative expansion,\footnote{The spin-dependent interactions in $H$ lead to admixtures in the wave function of components with $L_{ik}>0$, $L_{ik;j}>0$ \cite{isgur,phuoc-diss}. Because of the derivatives involved in the orbital angular momenta, the effects of these components can only appear explicitly through the derivative expansion.} this structure will be just that encompassed in the naive $L=0$ quark model, a result which will only be changed perturbatively by the effects of small spin-dependent interactions originating in QCD, and by quark mass effects. This suggests that effective field theory is essentially equivalent to the QM to order $m_s$, an observation we will explore in detail in the following sections. We will, in fact, reproduce the results of the nonrelativistic QM for baryon masses and moments\cite{halzen} in a completely relativistic context.

\subsection{Baryon masses in the symmetrical limit}
\label{subsec:baryon-masses}

The mass terms in the effective baryon Lagrangian are of the standard form
\begin{equation}
\label{L_0M}
{\cal L}_{0,M} = -m_B\,\bar{B}_{kji}B_{ijk} + m_T\,\bar{T}^\mu_{kji} T_{\mu;ijk}
\end{equation}
in the symmetrical limit $m_s=0$. For the semirelativistic Hamiltonian $H$ in Eq.\ (\ref{H_brambilla}), $m_B=<B|H|B>$ and $m_T=<T|H|T>$, so these masses involve contributions from the kinetic and potential energies, including the spin-dependent terms. Treating the latter as perturbations, the first-order difference of the energies in the $L=0$ ground state is given simply by the spin-spin term in Eq.\ (\ref{H_brambilla}), but there are generally also higher order contributions from the other spin-dependent interactions. 

The spin-spin structure is also embedded in the chiral description. To see this, we write the original mass terms in the effective Lagrangian as
\begin{equation}
\label{L_M2}
{\cal L}_M = -\bar{B}_{kji}\,m\,B_{ijk} + \bar{T}^\mu_{kji}\,m\, T_{\mu;ijk}
\end{equation}
where the total mass operator $m$ is given in terms of the separate masses and the spin-1/2 and spin-3/2 projection operators in Eqs.\ (\ref{project_1/2}) and (\ref{project_3/2}) by
\begin{eqnarray}
\label{BTmasses_QCD}
m &=& m_B\,P^{1/2}+m_T\,P^{3/2} \nonumber \\
&=& {\textstyle\frac{1}{2}}\,(m_T+m_B) + {\textstyle\frac{1}{6}}\,(m_T-m_B)\, \left(\mbox{\boldmath$\sigma$}_i\!\cdot\!\mbox{\boldmath$\sigma$}_j + \mbox{\boldmath$\sigma$}_j\!\cdot\!\mbox{\boldmath$\sigma$}_k +\mbox{\boldmath$\sigma$}_k\!\cdot\!\mbox{\boldmath$\sigma$}\right)_i.
\end{eqnarray}
We will identify the common mass $(m_T+m_B)/2$ of the octet and decuplet with the mass $m_0$ extracted in defining the heavy-baryon fields in Eq.\ (\ref{B_v}).  With this definition, $\delta m_T$ and $\delta m_B$ in Eq.\ (\ref{L0_B,T}) are simply $\pm\delta m$ where $\delta m = (m_T-m_B)/2$.  

The mass term ${\cal L}_M$ with $m_0$ removed reduces to
\begin{equation}
\label{L_M3}
{\cal L}_M = -\bar{B}_{kji}\,\Delta m\,B_{ijk} + \bar{T}^\mu_{kji}\,\Delta m\, T_{\mu;ijk} 
\end{equation}
where $\Delta m$ is the operator
\begin{equation}
\label{spin-mass-diff}
\Delta m = {\textstyle\frac{1}{3}}\,\delta m\, \left(\mbox{\boldmath$\sigma$}_i\!\cdot\!\mbox{\boldmath$\sigma$}_j + \mbox{\boldmath$\sigma$}_j\!\cdot\!\mbox{\boldmath$\sigma$}_k +\mbox{\boldmath$\sigma$}_k\!\cdot\!\mbox{\boldmath$\sigma$}\right)_i.
\end{equation}
This operator has the expected form of a spin-spin interaction, and has the values $\pm \delta m$ in the decuplet and octet. It is the only chiral invariant that contributes to the octet-decuplet mass difference in the symmetrical limit. 

The decuplet-octet mass difference is purely a QCD effect, ascribed in the semirelativistic Hamiltonian to the spin-spin interaction associated with short-distance gluon exchange. The relation in Eq.\ (\ref{spin-mass-diff}) is purely kinematic, and includes more than just the first-order spin-spin energy. However, the dynamical calculations of Capstick and Isgur \cite{isgur} show that a treatment of the mass difference in first-order perturbation theory is at least roughly correct, so $\delta m$ in fact gives a measure of the strength of the spin-spin interaction and the other spin-dependent terms in Eq.\ (\ref{H_brambilla}). It is fairly weak on the scale of the terms that determine the total masses. We will therefore assume that all the spin-dependent interactions are small and can be treated perturbatively, an assumption consistent with the dynamical calculations in \cite{kogut,isgur} and \cite{phuoc-diss}.

\subsection{Quark-level meson-baryon couplings}
\label{subsec:meson-baryon_couplings}

\subsubsection{Octet-octet-meson couplings}
\label{subsubsec:octet-octet-meson}

As a first example of a calculation in the three-flavor-index notation, we will explore the connection of the strong couplings $D$, $F$, $\cal C$, and $\cal H$   to the axial vector interaction
\begin{equation}
\label{qbarqA}
{\cal L}_A = \bar{q}_i^a\not\!\gamma^\mu\gamma^5 A_\mu q_i^a
\end{equation}
in the underlying quark-level chiral Lagrangian, where $A_\mu$ is the axial current defined in Eq.\ (\ref{Acurrent}). We suppose that there are no quark mass splittings so the theory is completely symmetric in the different quarks. We will use this calculation to develop methods we will need later.

Because the quark flavor lines are continuous through a diagram and the effective fields $B_{ijk}^{\,\gamma}$ and $T_{ijk}^{\,\mu\gamma}$ completely specify the flavor and spin structure of the external baryons, the structure of most general spin-flavor matrix elements for two baryons coupled through ${\cal L}_A$ can be determined using the explicit quark-level representations of the fields in Eqs.\ (\ref{Bijk}) and (\ref{Tijk}) and the dynamically allowed spin structures. The simplest spin structure is that implied by Eq.\ (\ref{qbarqA}), in which the spin on a quark line changes only because of the coupling to the axial current. This structure can be changed in the symmetrical limit by spin-dependent interactions within the baryons as discussed in connection with the baryon magnetic moments in Sec.\ \ref{sec:moments}. We expect the effective two-body operators introduced by these interactions to be small, and will not consider them here. 

The dynamical parts of the matrix elements can only be calculated using the underlying theory, but appear simply as unknown constants multiplying the independent spin-flavor matrix elements. Again, in the symmetrical limit, these constants will be equal for all baryons up to the perturbative effects of the spin-dependent interactions between quarks.

We will calculate the matrix elements of the quark-level axial interaction in Eq.\ (\ref{qbarqA}) in the baryon rest frame where $\gamma^\mu\gamma^5 A_\mu\rightarrow (0,\,-\mbox{\boldmath$\sigma$}\!\cdot {\bf A})$.\footnote{It is possible to carry out the calculation completely in covariant notation, as we have also done. However, the ostensibly noncovariant treatment above is considerably simpler and and more transparent with respect to the operations involving projections and quark interchanges. The final result is covariant.} Treating the quarks as non-interacting and suppressing the color indices, the spin-flavor matrix element for the octet baryon--meson interaction is then
\begin{eqnarray}
\label{bbM}
\left<\gamma' k'j'i'|\,(- \bar{q}_p\,\mbox{\boldmath$\sigma$}\!\cdot {\bf A}_{pp'}\, q_{p'})\,|ijk\gamma\right> &=& \left<0\,|B_{i'j'k'}^{\,\gamma'}\,(- \bar{q}_p\,\mbox{\boldmath$\sigma$}\!\cdot {\bf A}_{pp'}\, q_{p'})\, \bar{B}_{nml}^{\,\lambda} |\,0\right>P^{B;\lambda\gamma}_{lmn;kji} \nonumber \\ &=&-\left<0\,|B_{i'j'k'}^{\,\gamma'}\, {\textstyle\frac{1}{\sqrt{6}}}\left[(\bar{q}_p \,\mbox{\boldmath$\sigma$})^\lambda \!\cdot {\bf A}_{pn}\,\left(\bar{q}_{m} i\sigma_2\bar{q}_{l}^{\rm T}\right) \right.\right. \nonumber \\ && + \bar{q}_{n}^{\,\lambda}\, \left(\bar{q}_p\, i\mbox{\boldmath$\sigma$}\sigma_2\, \bar{q}_{l}^{\rm T}\right)\!\cdot\! {\bf A}_{pm}  \\ && \left.\left.- \bar{q}_{n}^{\,\lambda}\, \left(\bar{q}_p\, i\mbox{\boldmath$\sigma$}\sigma_2\, \bar{q}_{m}^{\rm T}\right)\!\cdot\!{\bf A}_{pl}\right]|\,0\right> P^{B;\lambda\gamma}_{lmn;kji}. \nonumber
\end{eqnarray}
We have inserted an apparently unnecessary rest-frame octet projection operator $P^{B;\lambda\gamma}_{lmn;kji}=\delta_{\lambda\gamma}P^B_{lmn;kji}$ for later convenience. 

The first term in the factor in square brackets has the spin operator $\mbox{\boldmath$\sigma$}$ acting on the odd quark, with the other two quarks in a singlet spin state. This involves the pure octet structure $\bar{B}_{pml}^{\,\gamma'}\,
(\mbox{\boldmath$\sigma$})_{\gamma'\lambda} \cdot{\bf A}_{pn}$. The remaining two terms have the paired quarks in a triplet spin state, and involve both octet and decuplet contributions. These can be isolated without calculating the final matrix element by using the projection operators $P^{1/2}$ and $P^{3/2}$ in Eqs.\ (\ref{project_1/2}) and (\ref{project_3/2}). Thus, acting on the second term with $P^{1/2}(lnp)$, we find that
\begin{eqnarray}
\label{P^1/2-B}
P^{1/2}(lnp)\,{\textstyle\frac{1}{\sqrt{6}}}\bar{q}_{n}^{\,\lambda}\, \left(\bar{q}_p\, i\mbox{\boldmath$\sigma$}\sigma_2\, \bar{q}_{l}^{\rm T}\right)\!\cdot\! {\bf A}_{lm} &=& \left[{\textstyle\frac{1}{6}}(1 - \mbox{\boldmath$\sigma$}_p \!\cdot \mbox{\boldmath$ \sigma$}_{l}) + {\textstyle\frac{1}{6}}(1 -\mbox{\boldmath$\sigma$}_{l} \!\cdot \mbox{\boldmath$ \sigma$}_{n}) + {\textstyle\frac{1}{6}}(1 -\mbox{\boldmath$\sigma$}_{n} \!\cdot \mbox{\boldmath$ \sigma$}_p) \right] \nonumber \\ && \times {\textstyle\frac{1}{\sqrt{6}}} \bar{q}_{n}^{\,\lambda}\, \left(\bar{q}_p\, i\mbox{\boldmath$\sigma$}\sigma_2\, \bar{q}_{l}^{\rm T}\right)\!\cdot\! {\bf A}_{pm} \nonumber \\ &=&{\textstyle\frac{1}{3}}\left(\bar{B}_{pnl}^{\,\lambda'} + \bar{B}_{lnp}^{\,\lambda'}\right) \,(\mbox{\boldmath$\sigma$})_{\lambda'\lambda} \!\cdot {\bf A}_{pm},
\end{eqnarray}
where we have noted that $(1 - \mbox{\boldmath$\sigma$}_p \cdot \mbox{\boldmath$ \sigma$}_{l})$ annihilates the original triplet combination of the quarks $q_p$ and $q_{l}$, and have used the result in Eq.\ (\ref{Pik-singlet}) to evaluate the effects of the remaining terms in $P^{1/2} (lnp)$. Thus, the operator $(1 -\mbox{\boldmath$\sigma$}_l \!\cdot \mbox{\boldmath$ \sigma$}_n)$ regroups $q_l$ and $q_n$ into the singlet combination which appears in $B$ and leaves the Pauli matrix $\mbox{\boldmath$\sigma$}$ acting only on the odd quark, that is, directly on the overall spinor index of $B_{pnl}^{\,\lambda'}$. A similar result holds for the the action of $(1 -\mbox{\boldmath$\sigma$}_n \!\cdot \mbox{\boldmath$ \sigma$}_p)$.

Upon combining all terms we find that 
\begin{eqnarray}
\label{BBM1}
\left<\gamma' k'j'i'|- \bar{q}_p\mbox{\boldmath$\sigma$}\!\cdot {\bf A}_{pp'}\, q_{p'}|ijk\gamma\right> &=& -\left<0\,|B_{i'j'k'}^{\,\gamma'}\, \left[\bar{B}_{pml}^{\,\lambda'}{\bf A}_{pn} + {\textstyle\frac{1}{3}}\left(\bar{B}_{pnl}^{\,\lambda'} + \bar{B}_{lnp}^{\,\lambda'}\right) {\bf A}_{pm}\right.\right.\nonumber \\ && -\left.\left.{\textstyle\frac{1}{3}}\left(\bar{B}_{pnm}^{\,\lambda'} + \bar{B}_{mnp}^{\,\lambda'}\right) {\bf A}_{pl}\,\right] \cdot(\mbox{\boldmath$\sigma$})_{\lambda'\lambda}|\,0\right> P^{B;\lambda\gamma}_{lmn;kji} \\
\label{BBM2}
&=& \left[\,-P^{B;\gamma'\lambda'}_{i'j'k';pml} {\bf A}_{pn} -{\textstyle\frac{1}{3}} \left(P^{B;\gamma'\lambda'}_{i'j'k';pnl} + P^{B;\gamma'\lambda'}_{i'j'k';lnp}\right){\bf A}_{pm} \right. \nonumber\\ && + \left.{\textstyle\frac{1}{3}}\left(P^{B;\gamma'\lambda'}_{i'j'k';pnm} + P^{B;\gamma'\lambda'}_{i'j'k';mnp}\right)\, {\bf A}_{pl}\,\right]\! \cdot(\mbox{\boldmath$\sigma$})_{\lambda'\lambda}\, P^{B;\lambda\gamma}_{lmn;kji} ,
\end{eqnarray}
where $P^B$ is the octet projection operator in Eq.\ (\ref{projection_op_oct}). 
To convert this expression into an effective baryon-level operator, we multiply on the left and right by $\bar{B}_{k'j'i'}^{\,\gamma'}$ and $B_{ijk}^{\,\gamma}$, respectively, sum over the indices, and use the projection property of $P^B$ given in Eq.\ (\ref{Bprojection}). The resulting operator reproduces the matrix element above when used instead of the quark-level operator $\bar{q}_l\mbox{\boldmath$\sigma$}\!\cdot {\bf A}_{ll'}\, q_{l'}$.\footnote{Had we not written $\bar{B}_{kji}$ in Eq.\ (\ref{bbM}) as $\bar{B}_{nml}P^B_{lmn;kji}$, the equation analogous to (\ref{BBM2}) would contain only one projection operator rather than the two obtained when calculating matrix elements of ${\cal L}_{BBM}$, Eq.\ (\ref{L_BBM}) below, using the general relation $\left<k'j'i'|\bar{B}OB|ijk\right> = P^B_{i'j'k';n'm'l'} O_{l'm'n';nml}P^B_{lmn;kji}$.  It can be shown that the results are equivalent, though this is not immediately obvious.} Finally, multiplying by the unknown dynamical matrix element $\beta$ at the vertex and relabeling indices, we obtain the effective baryon-level interaction
\begin{eqnarray}
\label{L_BBM}
{\cal L}_{BBM} &=& -\beta\, \bar{B}_{k'ji}\,\mbox{\boldmath$\sigma$} \!\cdot \!{\bf A}_{k'k} B_{ijk} \nonumber \\ && - {\textstyle\frac{1}{3}}\beta\,\left(\bar{B}_{j'ki} + \bar{B}_{ikj'}\right) \mbox{\boldmath$\sigma$}\,\!\cdot \!{\bf A}_{j'j} B_{ijk} \\ && + {\textstyle\frac{1}{3}}\beta\, \left(\bar{B}_{i'kj} + \bar{B}_{jki'}\right)\, \mbox{\boldmath$\sigma$} \!\cdot\! {\bf A}_{i'i}B_{ijk} \nonumber
\end{eqnarray}
where $\mbox{\boldmath$\sigma$}$ now appears in a spinor product. The last two terms can be combined if desired by using the symmmetries of the $B$'s and relabeling the summation indices. 

To connect ${\cal L}_{BBM}$ to the standard matrix form of HBChPT, we use the  Jacobi-like identity in Eq.\ (\ref{Jacobi}) and the identity\footnote{To prove this identity, we rewrite the product $\bar{B}_{i'jk}B_{ijk}$ as $\bar{B}_{i'jk}(-B_{jki}-B_{kij})= \bar{B}_{i'jk} B_{kji} -\bar{B}_{i'kj} B_{ikj}$ by using Eq.\ (\ref{Jacobi}) and the symmetries of the $B$'s. After a relabeling of the summed indices, the last term is identical to the left hand side up to its sign, and the result follows.}   
\begin{equation}
\label{BijkBijk}
\bar{B}_{i'jk}B_{ijk} = {\textstyle\frac{1}{2}}\bar{B}_{i'jk}B_{kji}
\end{equation}
to  rewrite the last term in Eq.\ (\ref{L_BBM}) as 
\begin{equation}
(\bar{B}_{i'kj}+\bar{B}_{jki'})\,B_{ijk} = -2\bar{B}_{i'jk}\,B_{ijk} + \bar{B}_{kji'}\,B_{ijk} = -\bar{B}_{i'jk}\,B_{kji}+\bar{B}_{kji'}\,B_{ijk}, 
\end{equation}
a form to which Eqs.\ (\ref{BbarBA}) and (\ref{BbarAB}) are applicable. The next-to-last term is treated similarly. The final result is 
\begin{equation}
\label{L_BBM-ChPT}
{\cal L}_{BBM} = 2\beta\,\left[\,{\rm Tr}\,\bar{B}\{\, S^\mu\! A_\mu,B\,\} + {\textstyle\frac{2}{3}}\,\bar{B}[\,S^\mu\! A_\mu,B\, ]\,\right],
\end{equation}
where we have used the correspondence $-\mbox{\boldmath$\sigma$}\! \cdot \!{\bf A}\equiv 2S^\mu\! A_\mu$ to put the expression in the standard covariant form. Comparing Eqs.\ (\ref{L_BBM-ChPT}) and (\ref{lag}), we see that the quark-level description gives the specific values
\begin{equation}
D=\beta,\qquad F={\textstyle\frac{2}{3}}\,\beta,
\end{equation}
for the couplings $D$ and $F$. These couplings automatically have the SU(6) ratio $F/D=2/3$.

We can also rewrite ${\cal L}_{BBM}$ in the expected form of a sum of one-body operators,
\begin{equation}
\label{L_BBM-one-body}
{\cal L}_{BBM} = -\beta\, \left[\bar{B}_{k'ji}\,\left(\mbox{\boldmath$\sigma$}\! \cdot {\bf A}\right)_{k'k}\,B_{ijk} + \bar{B}_{kj'i}\,\left(\mbox{\boldmath$\sigma$}\! \cdot {\bf A}\right)_{j'j}\,B_{ijk} + \bar{B}_{kji'}\,\left(\mbox{\boldmath$\sigma$}\! \cdot {\bf A}\right)_{i'i}\,B_{ijk}\right] ,  
\end{equation}
where the single-quark spin operator $\mbox{\boldmath$\sigma$}$ can be taken to act on either the final or initial quark. The first alternative leads directly to the expression for ${\cal L}_{BBM}$ in Eq.\ (\ref{L_BBM}) when evaluated as above. The second gives an expression related to the first by the symmetries of the $B$'s. 

${\cal L}_{BBM}$ is similarly given in covariant form by\footnote{ The structure for ${\cal L}_{BBM}$ in Eq.\ (\ref{L_BBM_rel}) can be obtained directly in a standard chiral calculation at the baryon level by determining how the original kinetic term $i\bar{B}\!\not\!\!\partial B$ for the effective fields changes when the effects of the Goldstone bosons are removed \cite{Weinberg_Vol_2}. The relevant transformation property of the $B$'s is given in Eq.\ (\ref{baryon_transform}), with $U$ now a spacetime-dependent $\gamma^5$ transformation. A similar calculation determines the form of ${\cal L}_{TTM}$ in Eq.\ (\ref{L_TTM_rel}) starting from the kinetic term $-i\bar{T}^\mu\!\not\!\!\partial T_\mu$. This procedure hides the connection of the results to the quark-level axial current $A_\mu$, and fails for ${\cal L}_{TBM}$, Eq.\ (\ref{L_TBM}), since there is no mixed $T,\,B$ kinetic term. While the result above is not surprising,  manipulations of the type used to obtain it are needed more generally to actually evaluate matrix elements in the three-flavor-index notation.} 
\begin{equation}
\label{L_BBM_rel}
{\cal L}_{BBM} = 2\beta\, \left[\bar{B}_{k'ji}\left(S^\mu\! A_\mu\right)_{k'k} B_{ijk} + \bar{B}_{kj'i}\left(S^\mu\! A_\mu \right)_{j'j}B_{ijk} + \bar{B}_{kji'}\left(S^\mu\! A_\mu\right)_{i'i}B_{ijk}\right],
\end{equation}
an expression that makes it clear that we are dealing with a relativistic effective field theory. The basic structure of ${\cal L}_{BBM}$ as a  symmetrical sum of individual quark-quark-meson interactions is also clear here, but is not clear in the usual expression in Eq.\ (\ref{L_BBM-ChPT}). This is a distinct advantage of the three-flavor-index representation for the fields.

\addtocounter{footnote}{-1}

We conclude this section by noting that spin exchange interactions within the baryons change the correlations between the initial and final quark spins and introduce effective couplings in which the operator $\left(\mbox{\boldmath$\sigma$}_{k}\! \cdot {\bf A}\right)_{k'k}$ in Eq.\ (\ref{L_BBM-one-body}) is replaced by $(\mbox{\boldmath$\sigma$}_{i}+\mbox{\boldmath$\sigma$}_{j})\! \cdot {\bf A}_{k'k}$, and similarly for the other terms. Matrix elements calculated with these operators have $F/D$ ratios different from 2/3 and therefore change the overall $F/D$ ratio, but by an amount we would expect to be small. 

\subsubsection{Octet-decuplet and decuplet-decuplet couplings}

We can obtain the octet-decuplet interaction in a similar fashion starting from the expression in Eq.\ (\ref{bbM}). The first term has a pure octet structure so does not connect to final decuplet states.  We can extract the decuplet components of the remaining two terms using the spin-3/2 projection operator $P^{3/2}$. Thus, using the relation in Eq.\ (\ref{Tproj}), 
\begin{eqnarray}
P^{3/2}{\textstyle\frac{1}{\sqrt{6}}}\left[\,\bar{q}_n^{\,\lambda}\, \left(\bar{q}_p\, i\mbox{\boldmath$\sigma$}\sigma_2\, \bar{q}_l^{\rm T}\right)\!\cdot\! {\bf A}_{pm} - \bar{q}_n^{\,\lambda}\, \left(\bar{q}_p\, i\mbox{\boldmath$\sigma$}\sigma_2\, \bar{q}_m^{\rm T}\right)\!\cdot\!{\bf A}_{pl}\,\right] = \sqrt{2}\,\left({\bf T}^{\,\lambda}_{nmp}\!\cdot\! {\bf A}_{pl} - {\bf T}^{\,\lambda}_{npl}\!\cdot \!{\bf A}_{pl}\right)
\end{eqnarray}
and we find that
\begin{equation}
\left<0|\,\bar{\bf T}^{\,\gamma'}_{k'j'i'}\, \left(-\bar{q}_p\, \mbox{\boldmath$\sigma$}\!\cdot {\bf A}_{pp'}q_{p'}\right)\, B^{\,\gamma}_{ijk}\,|0\right> = -2\sqrt{2}\,P^T_{i'j'k';nml'} {\bf A}_{l'l} P^B_{lmn;kji}\delta_{\gamma'\gamma},
\end{equation}
where we have used the symmetries of the projection operators to combine terms. 
Taking a scalar product with $\bar{\bf T}^{\,\gamma'}_{k'j'i'}$ on the left, multiplying on the right by $B_{ijk}^{\,\gamma}$, and summing over the indices using the properties of the projection operators, we obtain the baryon-level effective coupling
\begin{eqnarray}
\label{L_TBM}
{\cal L}_{TBM} &=& -2\sqrt{2}\beta'\,\left( \bar{\bf T}^{\,\gamma}_{kji'}\!\cdot \!{\bf A}_{i'i}\, B^{\,\gamma}_{ijk} + \bar{B}^{\,\gamma}_{kji'}\,{\bf A}_{i'i}\!\cdot\! {\bf T}^{\,\gamma}_{ijk}\,\right) \nonumber \\
&=& 2\sqrt{2}\beta'\,\left(\bar{T}^{\,\mu\gamma}_{kji'}\,A_{\mu,i'i}\,B^{\,\gamma}_{ijk} + \bar{B}^{\,\gamma}_{kji'}\,A_{\mu;i'i}\,T^{\,\mu\gamma}_{ijk}\right),
\end{eqnarray}
where we have inserted the unknown dynamical matrix element $\beta'$ and added the $T\rightarrow B$ terms. This is of the form of the standard coupling given in Eq.\ (\ref{L0_B,T}) with ${\cal C}=-2\beta'$. As discussed above, we expect that $\beta'=\beta$ up to the small corrections induced by the spin-dependent interactions between quarks, a result that goes beyond standard chiral symmetry arguments.\footnotemark 

We can obtain the effective decuplet-decuplet-meson interaction by a similar calculation, but with some further subtleties. We begin with the decuplet matrix element of the quark-level axial current  evaluated in the baryon rest frame, $\bar{q}_p \not\!\!\!\! A_{pp'}\gamma^5q_{p'} \rightarrow\left(0,-\bar{q}_p \,\mbox{\boldmath$\sigma$} \!\cdot\! {\bf A}_{pp'}q_{p'}\right)$. Inserting a factor of the decuplet projection operator of Eq.\ (\ref{projection_op_dec}) for convenience and treating the quarks as free, we obtain
\begin{eqnarray}
-\left<0\,|\,T^{r'\gamma'}_{i'j'k'}\,\bar{q}_p \,\mbox{\boldmath$\sigma$} \!\cdot\! {\bf A}_{pp'}q_{p'} \,\bar{T}^{r\gamma}_{kji}\,|\,0\right> &=& \left<0\,|\,T^{r'\gamma'}_{i'j'k'}\,{\textstyle\frac{1}{\!6\sqrt{3}}} \left[ \left(\bar{q}_p\, \mbox{\boldmath$\sigma$} \!\cdot\! {\bf A}_{pn}\right)^\lambda\, \left(\bar{q}_m\sigma_si\sigma_2 \bar{q}_l^{\,\rm T}\right)\right.\right. \nonumber \\ && +\left.\left. \bar{q}_n^\lambda \left(\bar{q}_p\, \mbox{\boldmath$\sigma$} \!\cdot\! {\bf A}_{pm} \sigma_s i\sigma_2 \bar{q}_l^{\,\rm T}\right) \right.\right. \\ && + \left.\left. \bar{q}_n^\lambda \left(\bar{q}_p\, \mbox{\boldmath$\sigma$} \!\cdot\! {\bf A}_{pl} \sigma_s i\sigma_2 \bar{q}_m^{\,\rm T}\right)  + \cdots \right]|\,0\right> P^{T\,s\lambda;r\gamma}_{lmn;kji}\,,\nonumber 
\end{eqnarray}
where $T^{r\gamma}$ is the $r$ component of ${\bf T}^\gamma$. The terms in ellipsis have the same structure, with $n\leftrightarrow m$ and $m\leftrightarrow l$. Applying the spin-3/2 projection operator $P^{3/2}$ in Eq.\ (\ref{project_3/2}) to the factor in square brackets and rearranging terms, we find that
\begin{eqnarray}
\label{TTmatrix}
-\left<0\,|\,T^{r'\gamma'}_{i'j'k'}\,\bar{q}_p \mbox{\boldmath$\sigma$} \!\cdot\! {\bf A}_{pp'}q_{p'} \,\bar{T}^{r\gamma}_{kji}\,|\,0\right> &=&
\left<0\,|\,T^{r'\gamma'}_{i'j'k'}\,{\textstyle\frac{1}{\!6\sqrt{3}}} \left[ \left(\bar{q}_p\, \mbox{\boldmath$\sigma$} \!\cdot\! {\bf A}_{pn}\right)^\lambda\, \left(\bar{q}_m\sigma_si\sigma_2 \bar{q}_l^{\,\rm T}\right)\right.\right. \nonumber \\ && + \left.\left. \bar{q}_p^\lambda \left(\bar{q}_m\, \mbox{\boldmath$\sigma$} \!\cdot\! {\bf A}_{pn}\, \sigma_s i\sigma_2 \bar{q}_l^{\,\rm T}\right) \right.\right. \\ && + \left.\left. \bar{q}_p^\lambda \left(\bar{q}_l\, \mbox{\boldmath$\sigma$} \!\cdot\! {\bf A}_{pn}\, \sigma_s i\sigma_2 \bar{q}_m^{\,\rm T}\right)  + \cdots \right]|\,0\right> P^{T\,s\lambda;r\gamma}_{lmn;kji} \,,\nonumber 
\end{eqnarray}
where the terms in ellipsis now involve the interchanges  $p\leftrightarrow m$ and $p\leftrightarrow l$. This is simply the relation
\begin{eqnarray}
-\left<0\,|\,T^{r'\gamma'}_{i'j'k'}\,\bar{q}_p \mbox{\boldmath$\sigma$} \!\cdot\! {\bf A}_{pp'}q_{p'} \,\bar{T}^{r\gamma}_{kji}\,|\,0\right> &=& -\left<0\,|\,T^{r'\gamma'}_{i'j'k'}\,\bar{T}^{s\lambda}_{pml}\,\left(\mbox{\boldmath$\sigma$}_p + \mbox{\boldmath$\sigma$}_m + \mbox{\boldmath$\sigma$}_l \right) \!\cdot\! {\bf A}_{pn}\,|\,0\right> P^{T\,s\lambda;r\gamma}_{lmn;kji} \nonumber \\ &=& -\left<0\,|\,T^{r'\gamma'}_{i'j'k'}\, \bar{T}^{s\lambda}_{pml}\, 2{\bf J \!\cdot\! A}_{pn}\,|\,0\right> P^{T\,s\lambda;r\gamma}_{lmn;kji} \,,\nonumber
\end{eqnarray}
where $\bf J$ is the total angular momentum operator of the spin-3/2 system taken to act on $\bar{T}$, $\bar{T}{\bf J}\equiv (\overline{{\bf J}T})$.

We can evaluate this expression explicitly by using the relation $\mbox{\boldmath$\sigma$}\!\cdot\!{\bf A}\,\sigma_s=A_s + i\epsilon_{ss't} \sigma_{s'}A_t$ and combining the last two terms in Eq.\ (\ref{TTmatrix}). The singlet pieces proportional to $A_s$ cancel with the result that, including the permutations, 
\begin{eqnarray}
\label{TxT}
-\left<0\,|\,T^{r'\gamma'}_{i'j'k'}\,\bar{q}_p \mbox{\boldmath$\sigma$} \!\cdot\! {\bf A}_{pp'}q_{p'} \,\bar{T}^{r\gamma}_{kji}\,|\,0\right>P^{T\,s\lambda;r\gamma}_{lmn;kji} &=& -\left<0\,|\,T^{r'\gamma'}_{i'j'k'}\,\bar{T}^{s'\lambda'}_{pml}\, \left[ \left(\mbox{ \boldmath$\sigma$} \right)_{\lambda'\lambda}\!\cdot\!{\bf A}_{pn}\delta_{s's}\right.\right.  \nonumber \\ && \left.\left. + i\epsilon_{ss't}\delta_{\lambda'\lambda}\,A^t_{pn}\right]\,|\,0\right> P^{T\,s\lambda;r\gamma}_{lmn;kji} \\ &=& 
-P^{T\,r'\gamma';s'\lambda'}_{i'j'k';pml} \left(\mbox{\boldmath$\sigma$} \!\cdot\!{\bf A}_{pn}\delta_{s's} +  i\epsilon_{ss't} \,A^t_{pn} \right) P^{T\,s\lambda;r\gamma}_{lmn;kji}. \nonumber
\end{eqnarray}
The effective interaction
\begin{eqnarray}
\label{L_TTM}
{\cal L}_{TTM}&=&-\beta''\,\bar{T}^{r'\gamma'}_{k'ji}\left( \mbox{\boldmath$\sigma$} \!\cdot\!{\bf A}_{pn}\delta_{r'r} +  i\epsilon_{rr't} \,A^t_{k'k}\delta_{\gamma'\gamma} \right)_{\gamma'\gamma} T^{r\gamma}_{ijk} \nonumber \\
&=&-\beta''\,\bar{\bf T}_{k'ji}\cdot (\mbox{\boldmath$\sigma$} \!\cdot\!{\bf A}_{k'k})\,{\bf T}_{ijk} + i\bar{\bf T}_{k'ji}\cdot {\bf A}_{k'k}\times{\bf T}_{ijk} \\
&=& -\beta''\,\bar{\bf T}_{k'ji}\cdot (2{\bf J\cdot A})_{k'k}{\bf T}_{ijk}
\end{eqnarray}
gives the same kinematic structure, with $\beta''\approx \beta$ the dynamical matrix element. The action of $\bf J$ on a vector-spinor operator has the standard definition $\hat{\epsilon}\cdot{\bf S} + \hat{\epsilon}\times$ where $\bf S = \frac{1}{2}\mbox{\boldmath$\sigma$}$ acts only on the spinor index.

In covariant form, 
\begin{equation}
\label{L_TTM_rel}
{\cal L}_{TTM} = -2\beta''\bar{T}_\mu J^\nu A_\nu T^\mu = -\bar{T}_\mu S^\nu A_\nu T^\mu - i\epsilon^{\sigma\mu\nu\lambda}v_\sigma \bar{T}_\mu A_\nu T_\lambda.
\end{equation}
\addtocounter{footnote}{-1}
We can also obtain this result directly from standard chiral field theory arguments with the $T$'s treated as elementary fields with the transformation properties in Eq.\ (\ref{delta_transform}),\footnotemark\  but with the loss of the simple connection to dynamical models. 

The decuplet-decuplet-meson coupling is usually stated in terms if $S^\nu A_\nu$ instead of $J^\nu A_\nu$ as in Eq.\ (\ref{L0_B,T}),
\begin{equation}
\label{L_TTM_spin}
{\cal L}_{TTM} = 2{\cal H}\,\bar{T}_\mu  S^\nu A_\nu T^\mu.
\end{equation} 
The two forms are connected by the Wigner-Eckart theorem applied to the corresponding rest-frame expressions. In particular, $\left<{\bf S\cdot A}\right>={\textstyle\frac{1}{3}}\left<{\bf J\cdot A}\right>$ for angular-\-momentum-\-3/2 states, a result that can be checked directly but does not arise from an obvious identity for the fields. Using this result and comparing Eqs.\ (\ref{L_TTM_rel}) and (\ref{L_TTM_spin}), we see that they are equivalent, with ${\cal H}=-3\beta''$. 

If spin-spin interactions are small as in the dynamical models discussed in Sec.\ \ref{subsec:dynamics},  $\beta'' \approx \beta' \approx \beta$ in the symmetrical limit, extra two-body contributions of the type discussed in Sec.\ \ref{sec:moments} are also small, and the results in Eqs. (\ref{L_BBM_rel}), (\ref{L_TBM}), and (\ref{L_TTM_rel}) reproduce the complete set of SU(6) coupling ratios for ${\cal L}_0$, Eq.\ (\ref{L0_B,T}),
\begin{equation}
\label{SU6_couplings}
F={\textstyle\frac{2}{3}D},\quad {\cal C}=-2D,\quad {\cal H}=-3D.
\end{equation}
Here $D=\beta$ is the common dynamical matrix element. Given the smallness of the spin-dependent interactions, the validity of the SU(6) coupling ratios for the ground-state baryons becomes a dynamical prediction rather than an input assumption. The dynamical matrix elements $\beta$, $\beta'$ and $\beta''$ will actually differ somewhat because of the small differences in the octet and decuplet wave functions induced by the spin-spin interactions, an effect which exists even in the symmetrical limit of QCD, and will also change in O($m_s$) because of symmetry-breaking quark mass effects. The coupling ratios will be further upset by the small contributions of two- and three-body operators. We will not consider these changes here.

\section{CHIRAL SYMMETRY BREAKING: BARYON MASSES AT O($m_s$)}
\label{sec:masses}

\subsection{One-body operators: O($m_s$) mass insertions}
\label{subsec:mass_insertions}

At the quark level, the symmetry-breaking mass terms in the chiral Lagrangian are of the one-body form
\begin{equation}
\label{quark_mass_terms}
{\cal L}_m = -m_s \,\bar{q}_l^{\,a}({\cal M}^+ +{\cal M}^-\gamma^5)_{ll'} q_{l'}^{\,a} ,
\end{equation}
where repeated indices are to be summed. The flavor matrices ${\cal M}^\pm$ are defined as
\begin{equation}
\label{Mpm}
{\cal M}_{\,ll'}^\pm = {\textstyle\frac{1}{2}}(\,\xi^\dagger M \xi^\dagger \pm \xi M\xi\,)_{ll'},
\end{equation}
where $M$ is the diagonal matrix 
\begin{equation}
\label{Mdefinition}
M = {\rm diag}\,(0,\,0,\,1\,).
\end{equation}
The matrix elements of $\gamma^5$ vanish in the heavy-baryon limit, and the $\gamma^5$ term in Eq.\ (\ref{quark_mass_terms}) can therefore be dropped.

In the absence of spin-dependent interactions, we can again use the explicit free-quark representations of the fields in Eqs.\ (\ref{Bijk}) and (\ref{Tijk}) to determine the most general spin-flavor matrix element for baryons coupled through ${\cal L}_m$. We treat the meson fields in the factors $\xi$ and $\xi^\dagger$ as elementary. Thus, suppressing the color indices, the spin-flavor matrix element for the octet baryons is
\begin{eqnarray}
\label{Bmass}
-m_s\left<0|B_{i'j'k'}^{\,\gamma'}\, \bar{q}_p {\cal M}^+_{pp'} q_{p'}\, \bar{B}_{kji}^{\,\gamma} |0\right> &=& -m_s\left<0|B_{i'j'k'}^{\,\gamma'}\, \bar{q}_p {\cal M}^+_{pp'} q_{p'}\, \bar{B}_{nml}^{\,\lambda} |0\right>P^{B;\lambda\gamma}_{lmn;kji} \nonumber \\ &=& -m_s\left(P^{B;\gamma'\lambda}_{i'j'k';n'ml}\, {\cal M}^+_{n'n}\,P^{B;\lambda\gamma}_{lmn;kji} \right. \\ && \left.  + P^{B;\gamma'\lambda}_{i'j'k';nm'l}\, {\cal M}^+_{l'l}\,P^{B;\lambda\gamma}_{lmn;kji}+P^{B;\gamma'\lambda}_{i'j'k';nml'} \,{\cal M}^+_{l'l}\, P^{B;\lambda\gamma}_{lmn;kji}\right)\,, \nonumber
\end{eqnarray}
where we have inserted an extra octet projection operator $P^B$, Eq.\ (\ref{projection_op_oct}), in the initial matrix element for symmetry. Multiplying as before on the left and right by $\bar{B}_{k'j'i'}^{\,\gamma'}$ and $B_{ijk}^{\,\gamma}$, respectively, summing over the indices, and using the projection property of $P^B$ given in Eq.\ (\ref{Bprojection}), we obtain an effective operator that reproduces the matrix element above when used instead of the quark-level operator $\bar{q}_l{\cal M}^+_{ll'} q_{l'}$, 
\begin{equation}
\label{L^B_m}
{\cal L}^B_m = -\alpha_m\,\left(\bar{B}_{k'ji}^{\,\gamma}\,{\cal M}^+_{k'k}\,B_{ijk}^{\,\gamma} + \bar{B}_{kj'i}^{\,\gamma}\,{\cal M}^+_{j'j}\,B_{ijk}^{\,\gamma} + \bar{B}_{kji'}^{\,\gamma}\,{\cal M}^+_{i'i}\,B_{ijk}^{\,\gamma}\right)\,.
\end{equation}
Here $\alpha_m$ is the unknown dynamical matrix element, with the factor of $m_s$ absorbed. This effective interaction has an obvious one-body structure.

After a similar calculation, we find the effective O($m_s$) one-body mass operator for the decuplet baryons,
\begin{eqnarray}
\label{L^T_m}
{\cal L}^T_m &=& \alpha'_m\,\left(\bar{T}_{k'ji}^{\,\mu\gamma}\,{\cal M}^+_{k'k}\,T_{\mu;ijk}^{\,\gamma} + \bar{T}_{kj'i}^{\,\mu\gamma}\,{\cal M}^+_{j'j}\,T_{\mu;ijk}^{\,\gamma} + \bar{T}_{kji'}^{\,\mu\gamma}\,{\cal M}^+_{i'i}\,T_{\mu;ijk}^{\,\gamma}\right) \nonumber\\
&=& 3\alpha'_m\,\bar{T}_{k'ji}^{\,\mu\gamma}\,{\cal M}^+_{k'k}\,T_{\mu;ijk}^{\,\gamma}.
\end{eqnarray}

We could, of course, have written these operators down directly as possible O($m_s$) baryon-level chiral invariants without the calculations above. However, the structure of these operators follows directly from the the structure of the quark-level chiral mass term in Eq.\ (\ref{quark_mass_terms}) and the continuity of flavor lines through the actual dynamical process, assuming no spin exchange. The invariants above are in fact the only O($m_s$) mass invariants with the one-body structure. Furthermore, from the connection to dynamical models, we expect the octet and decuplet matrix elements $\alpha_m$ and $\alpha'_m$ to be equal in the symmetrical limit $m_s=0$ except for small spin effects. We will henceforth assume this equality.

The situation is much less clear if we rewrite the octet operator in Eq.\ (\ref{L^B_m}) in the standard matrix form using the definitions in Eq.\ (\ref{Bkl}) or the relations in Eqs.\ (\ref{BbarBA}) and (\ref{BbarAB}). This gives
\begin{equation}
\label{chiral_mass}
{\cal L}^B_m = -\alpha_m\,{\rm Tr}\,\bar{B}[{\cal M}^+,B] - \alpha_m\,({\rm Tr}\,{\cal M}^+)\,({\rm Tr}\,\bar{B}B)\,.
\end{equation}
The simple connection to the one-body structure is lost. Comparing this result with the most general O($m_s$) mass corrections to the heavy-baryon Lagrangian \cite{Jen-HBChPT,Jen-masses},
\begin{eqnarray}
\label{L_ms}
{\cal L}_{m_s} &=& 2b_D\,{\rm Tr}\,\bar{B}\{{\cal M}^+,B\} + 2b_F\,{\rm Tr}\,\bar{B} [{\cal M}^+,B] + 2c\,\bar{T}^\mu{\cal M}^+T_\mu \nonumber \\
&& +2\sigma\, ({\rm Tr}\,{\cal M}^+) \,({\rm Tr}\,\bar{B}B) + 2\tilde{\sigma}\, {\rm Tr} \,({\cal M}^+)\,(\bar{T}^\mu T_\mu),
\end{eqnarray}
we see that the one-body mass insertions give a specific octet structure with an F-type mass term with $2b_F=-\alpha_m$, a related $\sigma$-type term with  $2\sigma=-\alpha_m$, and no D-type contribution. The decuplet terms have $2c=3\alpha'_m\approx 3\alpha_m$, and $\tilde{\sigma}=0$. The use of the quark-level picture with its connection to dynamics provides extra information. Any further contributions to ${\cal L}_{m_s}$ must come from two- or three-body operators.

\subsection{Two-body operators: O($m_s$) spin-spin interactions}
\label{subsec:spin-spin_terms}

The only two-body operators of order $m_s$ are of the spin-spin type as suggested by the semirelativistic dynamical model in Sec.\ \ref{subsec:dynamics}. In a chiral theory in the heavy baryon limit, the only quark-level spin-spin operator has the form
\begin{equation}
\label{spin-spin-Dirac}
O_{ss} = {\textstyle\frac{1}{2}} \sum_{p\not=r}\,:\!(\bar{q}_{p}\,\mbox{\boldmath$\sigma$}{\cal M}^+_{pp'}\,q_{p'})\cdot(\bar{q}_{r}\, \mbox{\boldmath$\sigma$}q_{r})\!:
\end{equation}
in the baryon rest frame, where, from Eq.\ (\ref{eq:Dirac_reduction}), the $\mbox{\boldmath$\sigma_n$}\cdot \mbox{\boldmath$\sigma_l$}$ structure can arise from either axial vector or Pauli couplings, $(\gamma^\mu\gamma^5)_n(\gamma_\mu\gamma^5)_{l}$ or $(\sigma^{\mu\nu})_n(\sigma_{\mu\nu})_{l}$. We treat the quark fields as normal-ordered to eliminate effective one-body operators. 

The matrix elements of this scalar operator do not connect octet and decuplet states. They reduce in octet baryon states, expressed in terms of the baryon fields, to
\begin{eqnarray}
\left<0\,|\,B_{i'j'k'}^{\,\gamma'}\, O_{ss}\, \bar{B}_{kji}^{\,\gamma}\, |\,0\right> &=& 
\left<0\,|\,B_{i'j'k'}^{\,\gamma'}\left[ \bar{B}_{nmp}^{\,\lambda}\,\mbox{\boldmath$\sigma$}_p\!\cdot (\mbox{\boldmath$\sigma$}_m+\mbox{\boldmath$\sigma$}_n) \, {\cal M}^+_{pl} + \bar{B}_{npl}^{\,\lambda}\,\mbox{\boldmath$\sigma$}_p\!\cdot (\mbox{\boldmath$\sigma$}_n+\mbox{\boldmath$\sigma$}_l)\,  {\cal M}^+_{pm} \right.\right. \nonumber \\ && \left.\left. + \bar{B}_{pml}^{\,\lambda}\,\mbox{\boldmath$\sigma$}_p\!\cdot (\mbox{\boldmath$\sigma$}_l +\mbox{\boldmath$\sigma$}_m)\,  {\cal M}^+_{pn}\,\right] |\,0\right>P^{B;\lambda\gamma}_{lmn;kji} .
\end{eqnarray}
The last term vanishes because quarks $l$ and $m$ are in a singlet spin state.   We can determine the action of the remaining spin operators on $\bar{B}$ by using their connection to the permutation operators $P_{nl}$ given in Eq.\ (\ref{Pij}), $\mbox{\boldmath$\sigma$}_p \!\cdot \! \mbox{\boldmath$\sigma$}_r = 2P_{rs}-1$. With this identification, 
\begin{eqnarray}
\label{BO_ssB}
\left<0\,|\,B_{i'j'k'}^{\,\gamma'}\, O_{ss}\, \bar{B}_{kji}^{\,\gamma}\, |\,0\right> &=& 
2\left<0\,|\,B_{i'j'k'}^{\,\gamma'}\, [\,(\bar{B}_{pmn}^{\,\lambda} + \bar{B}_{npm}^{\,\lambda} - \bar{B}_{nmp}^{\,\lambda} ) {\cal M}^+_{pl} - (l\leftrightarrow m) |\,0\right> P^{B;\lambda\gamma}_{lmn;kji} \nonumber \\ &=& -2\left<0\,|\,B_{i'j'k'}^{\,\gamma'}\, [\, (4\bar{B}_{nmp}^{\,\lambda}- 2\bar{B}_{pmn}^{\,\lambda}  ) {\cal M}^+_{pl}\,|\,0\right> P^{B;\lambda\gamma}_{lmn;kji}  \\ &=& -2\left(4P^{B;\gamma'\lambda}_{i'j'k';nmp} -2 P^{B;\gamma'\lambda}_{i'j'k';pmn}  \right) {\cal M}^+_{pl}\,  P^{B;\lambda\gamma}_{lmn;kji} , \nonumber
\end{eqnarray}
where we have used the antisymmetry of $\bar{B}_{nml}$ and $P^{B}_{lmn;kji}$
in $l,\,m$ and a relabeling of the summation indices to combine terms. 

The baryon-level effective interaction that yields this matrix element is
\begin{eqnarray}
\label{M_B_ss}
{\cal L}^B_{ss} &=& -2\alpha_{ss}\left(4\bar{B}_{kji'}{\cal M}^+_{i'i} B_{ijk} - 2\bar{B}_{i'jk}{\cal M}^+_{i'i} B_{ijk}\right) \nonumber \\ &=& -2\alpha_{ss}\left(4\bar{B}_{kji'}{\cal M}^+_{i'i}B_{ijk} - \bar{B}_{k'ji}{\cal M}^+_{k'k}B_{ijk}\right),
\end{eqnarray}
where we have used the relation in Eq.\ (\ref{BijkBijk}) and a relabeling of indices in writing the second line. $\alpha_{ss}$ is the O($m_s$) dynamical matrix element. In this form, we can use the relations in Eqs.\ (\ref{BbarBA}) and (\ref{BbarAB}) to write ${\cal L}^B_{ss}$ in matrix form, with the result
\begin{equation}
\label{M_ss_matrix}
{\cal L}^B_{ss} = 3\alpha_{ss}\,{\rm Tr}\,\bar{B}\{{\cal M}^+,B\} - \alpha_{ss}\,{\rm Tr}\,\bar{B}[{\cal M}^+,B] - 4\alpha_{ss}\,({\rm Tr}\,{\cal M}^+)({\rm Tr}\,\bar{B}B)\,.
\end{equation}
The first two terms have the standard form in Eq.\ (\ref{L_ms}), with
$2b_D=3\alpha_{ss}$ and $2b_F=-\alpha_{ss}$.  The final term is again a $\sigma$ term with a specified coefficient.

The contributions of the two-body spin-spin interaction to the decuplet mass operator are relatively simple. Each of the quark pairs is in a triplet spin configuration with $\mbox{\boldmath$\sigma$}_r \!\cdot \mbox{\boldmath$\sigma$}_s = 1$, and the fields $T_{lmn}$ are completely symmetric in the indices $l,\,m,\,n$. As a result, following the structure in Eq.\ (\ref{BO_ssB}), 
\begin{eqnarray}
\label{TO_ssT}
\left<0\,|\,T^{r'\gamma'}_{i'j'k'}\,O_{ss}\,\bar{T}^{r,\gamma}_{kji}\, |\,0\right> &=& 2\left<0\,|\,T^{r'\gamma'}_{i'j'k'}\,\bar{T}^{s\lambda}_{nmp}\, {\cal M}_{pl} + (l\leftrightarrow m) +(l\leftrightarrow n)\,|\,0\right> \nonumber \\ &=& 6\,P^{T\,r'\gamma';s\lambda}_{i'j'k';nmp}\,{\cal M}_{pl}\, P^{T\,s\lambda;r\gamma}_{lmn;kji}.
\end{eqnarray}
The corresponding effective interaction is
\begin{equation}
\label{M_T_ss}
{\cal L}^T_{ss}=6\alpha'_{ss}\,\bar{T}^r_{k'ji}\,{\cal M}_{k'k}\,T^r_{ijk} = -6\alpha'_{ss}\,\bar{T}^\mu_{k'ji}\,{\cal M}_{k'k}\,T_{\mu;ijk}.
\end{equation}
We expect that $\alpha'_{ss}\approx\alpha_{ss}$, a relation that would be an equality in leading order in the spin-spin interactions, but which is only approximate when higher-order effects are included \cite{isgur}.

The only three-body spin-spin interactions have the structures of $\alpha''_{ss}\mbox{\boldmath$\sigma$}_i\!\cdot\mbox{\boldmath$\sigma$}_j {\cal M}^+_{k'k}$ and its permutations. These terms arise from changes in the $i,\,j$ spin-spin matrix element caused by a non-zero mass correction for the third quark $k$. However, by adding and subtracting operators $\alpha''_{ss}(\mbox{\boldmath$\sigma$}_i + \mbox{\boldmath$\sigma$}_j)\!\cdot \mbox{\boldmath$\sigma$}_k {\cal M}^+_{k'k}$ of the type considered above, we obtain as the only new structure
\begin{equation}
\alpha''_{ss}\,(\mbox{\boldmath$\sigma$}_i\!\cdot\mbox{\boldmath$\sigma$}_j + \mbox{\boldmath$\sigma$}_j\!\cdot\mbox{\boldmath$\sigma$}_k + \mbox{\boldmath$\sigma$}_k\!\cdot\mbox{\boldmath$\sigma$}_i)\,{\cal M}^+_{k'k} +\cdots.
\end{equation}
The spin factor has the value $-3$ ($+3$) on the octet (decuplet) states, leaving just the structure of a one-body quark mass correction. Thus, including both contributions and taking $\alpha''_{ss}$ the same for the decuplet and octet, a reasonable approximation, the only effect of the $\alpha''_{ss}\mbox{\boldmath$\sigma$}_i\!\cdot\mbox{\boldmath$\sigma$}_j {\cal M}^+_{k'k}$ term is to change the effective values of $\alpha_m$ and $\alpha'_m$ in Eqs.\ (\ref{L^B_m}) and (\ref{L^T_m}), and of $\alpha_{ss}$ and $\alpha'_{ss}$ in Eq.\ (\ref{M_B_ss}) and (\ref{M_T_ss}) to 
\begin{eqnarray}
\label{alpha''_corr}
\alpha_m &\rightarrow& \tilde{\alpha}_m=\alpha_m+3\alpha''_{ss}, \qquad \alpha'_m \rightarrow \tilde{\alpha}'_m=\alpha'_m-3\alpha''_{ss}, \\ \alpha_{ss} &\rightarrow& \tilde{\alpha}_{ss}=\alpha_{ss}-\alpha''_{ss},\  \qquad \alpha'_{ss} \rightarrow \tilde{\alpha}'_{ss}=\alpha'_{ss}-\alpha''_{ss}.
\end{eqnarray}

\subsection{Baryon masses and the quark model}
\label{subsec:result_masses}

The complete expression for the baryon mass Lagrangian obtained by combining the terms in Secs.\ \ref{subsec:baryon-masses}, \ref{subsec:mass_insertions}, and \ref{subsec:spin-spin_terms} is
\begin{equation}
{\cal L}_M={\cal L}^B_M+{\cal L}^T_M
\end{equation}
where, to leading order in the derivative expansion of HBChPT and to O($m_s$), %
\begin{eqnarray}
\label{L^B_M}
{\cal L}^B_M &=& \delta m\,\bar{B}_{kji}\,B_{ijk} -\tilde{\alpha}_m\,\left(\bar{B}_{k'ji}^{\,\gamma}\,{\cal M}^+_{k'k}\,B_{ijk}^{\,\gamma} + \bar{B}_{kj'i}^{\,\gamma}\,{\cal M}^+_{j'j}\,B_{ijk}^{\,\gamma} + \bar{B}_{kji'}^{\,\gamma}\,{\cal M}^+_{i'i}\,B_{ijk}^{\,\gamma}\right) \nonumber \\ && -2 \tilde{\alpha}_{ss} \left(4\bar{B}_{kji'}^{\,\gamma}\,{\cal M}^+_{i'i}\,B_{ijk}^{\,\gamma}-\bar{B}_{k'ji}^{\,\gamma}\,{\cal M}^+_{k'k}\,B_{ijk}^{\,\gamma}\right) \\ \label{LB_TrM}
&=& [\,\delta m-(\tilde{\alpha}_m+4\tilde{\alpha}_{ss})\,{\rm Tr}{\cal M}^+\,]            \, {\rm Tr}\,\bar{B}B \nonumber \\ && - (\tilde{\alpha}_m+\tilde{\alpha}_{ss}) \,{\rm Tr}\,\bar{B}[{\cal M}^+,\,B] + 3\tilde{\alpha}_{ss}\,{\rm Tr}\,\bar{B}\{ {\cal M}^+,\,B\} 
\end{eqnarray}
and
\begin{equation}
\label{L^T_M}
{\cal L}^T_M = \delta m\,\bar{T}^\mu_{kji}\, T_{\mu;ijk} +  3(\tilde{\alpha}'_m-2\tilde{\alpha}'_{ss})\,\bar{T}_{k'ji}^{\,\mu\gamma}\,{\cal M}^+_{k'k}\,T_{\mu;ijk}^{\,\gamma}.
\end{equation}

We note that the term proportional to ${\rm Tr}{\cal M}^+$ in the matrix expression in Eq.\ (\ref{LB_TrM}) has the form of a octet mass term when taken  to leading order in the meson fields so that ${\rm Tr}{\cal M}^+\rightarrow1$. While this contribution can be eliminated by redefining $m_0$ and $\delta m$ in Eq.\ (\ref{BTmasses_QCD}),\footnote{In general, we can incorporate the vacuum expectation value of ${\rm Tr}{\cal M}^+$ with respect to the meson fields into $m_B$, redefine $m_0$ and $\delta m$ as
\begin{displaymath}
m_0\rightarrow m_0+{\textstyle\frac{1}{2}}(\alpha_m+4\tilde{\alpha}''_{ss})\left<0\,|{\rm Tr}{\cal M}^+|\,0\right>, \qquad \delta m\rightarrow \delta m - {\textstyle\frac{1}{2}}(\alpha_m+4\tilde{\alpha}''_{ss})\left<0\,|{\rm Tr}{\cal M}^+|\,0\right>,
\end{displaymath}
and replace ${\rm Tr}{\cal M}^+$ in Eq.\ (\ref{L^B_M}) by ${\rm Tr}{\cal M}^+ -\left<0\,|{\rm Tr}{\cal M}^+|\,0\right>$, a form that is O($\phi^2$) in the meson field.} as has been done implicitly in making numerical fits to hadron masses, for example, in \cite{DH_masses}, that procedure hides the simple connection of our results to the underlying quark structure. 

Comparing our expressions for ${\cal L}^B_M$ and ${\cal L}^T_M$ with the form of the O($m_s$) mass terms given in Eq.\ (\ref{L_ms}) \cite{Jen-HBChPT,Jen-masses}, we see that the two forms are completely equivalent, with
\begin{eqnarray}
\label{mass_parameters}
&&2b_F =-\alpha_m-\alpha_{ss}-2\alpha''_{ss}, \qquad \  2b_D = 3(\alpha_{ss}-\alpha''_{ss}), \nonumber \\ &&2c = 3\alpha'_m-6\alpha'_{ss}-3\alpha''_{ss}, \qquad 2\sigma= -(\alpha_m+4\alpha_{ss} - \alpha''_{ss}), \qquad 2\tilde{\sigma}=0.
\end{eqnarray}
However, the quark description has the advantage that the various contributions have direct physical interpretations in terms of the underlying dynamics. In particular, the $b_D$ and $\sigma$ mass terms arise entirely from spin-spin interactions or correlations, while the $b_F$ and $c$ terms also involve direct quark mass corrections to the main, spin-independent part of the energy. 

To see that this interpretation is reasonable as far as the sizes of the terms are concerned, we can evaluate the parameters using the results of a direct fit to the baryon masses. Assuming that the relations $\alpha'_m\approx\alpha_m$ and $\alpha'_{ss}\approx\alpha_{ss}$ can be treated as equalities, we find that $\delta m=146.5$ MeV, $\alpha_m=\alpha'_m = 178.4$ MeV, $\alpha_{ss}=\alpha'_{ss} = 17.1$ MeV, and $\alpha''_{ss} = -2.9$ MeV, with a mean deviation of the fit from the experimental masses of 3.1 MeV. The spin-independent mass corrections have the sign and general magnitude expected for the replacement of a light quark by a strange quark. The main two-body spin-spin term $\delta m$ has a similar magnitude, and the sign corresponding to the expected color spin-spin interaction, repulsive in the decuplet states. The mass correction $\alpha_{ss}$ to the spin-spin term is substantially smaller as is expected for a short-range QCD interaction of the type in Eq. (\ref{H_brambilla}), has the expected sign, and are also larger than the three-body term $\alpha''_{ss}$. These smaller terms are sensitive numerically to the validity of the approximation $\alpha_m=\alpha'_m$, but the relative magnitudes are stable. The individual contributions can only be separated completely using further dynamical input. 

The structure of these results is exactly that assumed in the nonrelativistic constituent quark model \cite{halzen}. The free, but off-shell, quarks used in the description of the $B$'s and $T$'s act in effect like nonrelativistic constituent quarks with small momenta inside the heavy baryon. However, we would reemphasize that we are actually dealing with a relativistic effective field theory in the heavy-baryon limit. The three-flavor-index ``quark'' representation of the baryons describes all the possible spin and flavor correlations in the relativistic matrix elements.  The internal momentum structure of the baryons only appears explicitly with higher terms in the momentum expansion. The spin correlations can be connected directly to the underlying dynamics when spin-dependent forces are weak, an important point for interpretation and applications as we will show in a subsequent paper on meson loop corrections.

\section{BARYON MAGNETIC MOMENTS AT O($m_s$)}
\label{sec:moments}

To complete our discussion of the connection between relativistic HBChPT and the quark model, we will sketch the parametrization of the baryon magnetic moments. The relevant calculational procedures have all been developed above. 

\subsection{One-body operators}
\label{subsec:moments_onebody}

The interaction Lagrangian for a quark magnetic moment in an external electromagnetic field is proportional to the operator
\begin{equation}
\label{mu_Q}
{\textstyle\frac{1}{2}}\,\bar{q}_{p} \,\sigma^{\lambda\nu}F_{\lambda\nu}\,Q_{pp'}\,q_{p'} , 
\end{equation}
where $Q$ is the diagonal quark charge matrix, $Q={\rm diag}(2/3,\,-1/3,\,-1/3)$. We can determine the matrix elements of this operator in the absence of multibody spin-dependent interactions using the results of Sec.\ \ref{subsec:meson-baryon_couplings}. Thus, working in the baryon rest frame where $\frac{1}{2}\,\sigma^{\lambda\nu}F_{\lambda\nu}\rightarrow (0, -\mbox{\boldmath$\sigma$}\!\cdot\!{\bf B})$ with $\bf B$ is the magnetic field,     
we find the same structure as in Sec.\ \ref{subsubsec:octet-octet-meson} with the axial current ${\bf A}$ replaced by $Q{\bf B}$. We therefore find from Eq.\ (\ref{L_BBM}) that the one-body octet moment interactions are given by the effective Lagrangian
\begin{eqnarray}
\label{mu0_BB}
{\cal L}^0_{\mu,BB} &=&\mu_1\, \left[\, \bar{B}_{k'ji}\,Q_{k'k} \,\mbox{\boldmath$\sigma$}\, B_{ijk}  \right. \nonumber \\ && - \left. {\textstyle\frac{1}{3}}\,\left(\bar{B}_{j'ki} + \bar{B}_{ikj'}\right) Q_{j'j}\,\mbox{\boldmath$\sigma$}\, B_{ijk}  +  {\textstyle\frac{1}{3}}\, \left(\bar{B}_{i'kj} + \bar{B}_{jki'}\right) Q_{i'i}\,\mbox{\boldmath$\sigma$} \,B_{ijk}\, \right]  \cdot  {\bf B} \\ &=& \mu_1\left(\, {\textstyle\frac{5}{3}} \,\bar{B}_{k'ji}\,Q_{k'k}\,\mbox{\boldmath$\sigma$}\, B_{ijk} - {\textstyle\frac{2}{3}}\,\bar{B}_{kj'i}\, Q_{j'j}\, \mbox{\boldmath$\sigma$}\, B_{ijk}\,\right)  \cdot  {\bf B}.\nonumber
\end{eqnarray}

The dynamical matrix element $\mu_1$ is changed in first order by the symmetry-breaking mass of the interacting quark through a second one-body operator
\begin{equation}
\label{mu_QM}
{\textstyle\frac{1}{2}}\,\bar{q}_{i'} \,\sigma^{\mu\nu}F_{\mu\nu}\,(QM)_{i'i}\,q_i,
\end{equation}
where $M={\rm diag}(0,\,0,\,1)$. Since $MQ=QM$, the operators in Eqs.\ (\ref{mu_Q}) and (\ref{mu_QM}) are the only one-body operators. These have the same spin structure and combine in the total matrix element to give the effective interaction
\begin{equation}
\label{mu_BB}
{\cal L}_{\mu,BB} = \mbox{\boldmath$\mu$}\!\cdot{\bf B} =\left({\textstyle\frac{5}{3}}\,\bar{B}_{k'ji}\,\mu_{k'k}\,\mbox{\boldmath$\sigma$}\, B_{ijk} - {\textstyle\frac{2}{3}}\,\bar{B}_{kj'i}\, \mu_{j'j}\, \mbox{\boldmath$\sigma$}\, B_{ijk}\right)\cdot{\bf B}.
\end{equation}
Here $\mu$ is the matrix
\begin{equation}
\label{mu_1,2}
\mu=\mu_1\,Q+\mu_2\,QM,
\end{equation}
and $\mbox{\boldmath$\mu$}$ is the effective baryon magnetic moment operator corresponding to the interaction Hamiltonian ${\cal H}_{\mu,BB}= -\mbox{\boldmath$\mu$}\cdot{\bf B}$. This structure generalizes in an arbitrary Lorentz frame to 
\begin{equation}
\label{mu_BB_rel}
{\cal L}_{\mu,BB} = -\left( \,{\textstyle\frac{5}{6}}\, \bar{B}_{k'ji}\,\mu_{k'k}\,\sigma^{\lambda\nu}\, B_{ijk} - {\textstyle\frac{1}{3}}\,\bar{B}_{kj'i}\, \mu_{j'j}\, \sigma^{\lambda\nu}\, B_{ijk}\,\right) \,F_{\lambda\nu}.
\end{equation}

We can put this ${\cal L}_{\mu,BB}$ in matrix form by using the relations in Eqs.\ (\ref{mu_1,2}), (\ref{BbarBA}) and (\ref{BbarAB}), with the result
\begin{eqnarray}
\label{mu_matrix}
{\cal L}_{\mu,BB} &=& \left(\,{\textstyle\frac{5}{6}}\, {\rm Tr}\,\bar{B}\,\mu\, \sigma^{\lambda\nu}B + {\textstyle\frac{1}{6}}\,{\rm Tr}\,\bar{B}\sigma^{\lambda\nu}B \mu - {\textstyle\frac{1}{6}}\,{\rm Tr}\,\mu\;{\rm Tr}\,\bar{B}\sigma^{\lambda\nu}B \,\right)F_{\lambda\nu}  \\ \label{one-body-moments} &=& \mu_1\,\left(\,{\textstyle\frac{5}{6}}\, {\rm Tr}\,\bar{B}\,Q\, \sigma^{\lambda\nu}B + {\textstyle\frac{1}{6}}\,{\rm Tr}\,\bar{B}\sigma^{\lambda\nu}B Q \right)F_{\lambda\nu} \nonumber \\   &&+\mu_2\,\left(\,
{\textstyle\frac{5}{6}}\, {\rm Tr}\,\bar{B}\,MQ\, \sigma^{\lambda\nu}B + {\textstyle\frac{1}{6}}\,{\rm Tr}\,\bar{B}\sigma^{\lambda\nu}B MQ - {\textstyle\frac{1}{6}}\,{\rm Tr}\,MQ\;{\rm Tr}\,\bar{B}\sigma^{\lambda\nu}B \, \right)F_{\lambda\nu},
\end{eqnarray}
where we have used ${\rm Tr}\,Q=0$. The line in this equation with the prefactor $\mu_1$ has the same form as the octet-octet-meson coupling in Sec.\ \ref{subsubsec:octet-octet-meson} and can be identified with the Coleman-Glashow form for the moment operator\cite{coleman}
\begin{equation}
\label{CGmoments}
{\cal L}_{CG} = \frac{e}{4m_N}\,\left(\mu_D\,{\rm Tr}\,\bar{B}\{Q,\sigma^{\lambda\nu}B\} + \mu_F\, {\rm Tr}\,\bar{B}[Q,\sigma^{\lambda\nu}B]\right)F_{\lambda\nu}
\end{equation}
for the SU(6)-symmetric choice of parameters $(e/2m_N)\mu_D=\mu_1$,  $(e/2m_N)\mu_F=2\mu_1/3$, or $\mu_F/\mu_D=2/3$. The remaining terms give the quark-mass corrections to the one-body moment operator. The complete expression in Eq.\ (\ref{mu_matrix}) is exactly that obtained in the simple additive quark model \cite{DH-loop-moments}. Two- and three-body effects lead to small deviations from this result as we will show later.

We can also obtain the one-body contributions to the decuplet-octet transition moments and the decuplet magnetic moments from the results in Sec.\ \ref{subsec:meson-baryon_couplings}. Thus, following the calculations that lead to Eq.\ (\ref{L_TBM}), we obtain the magnetic transition operator
\begin{eqnarray}
\label{L_muTB}
{\cal L}_{\mu,TB} &=& 2\sqrt{2}\,\left( \bar{\bf T}^{\,\gamma}_{kji'}\,\mu_{i'i}\, B^{\,\gamma}_{ijk} + \bar{B}^{\,\gamma}_{kji'} \,\mu_{i'i}\!\cdot\! {\bf T}^{\,\gamma}_{ijk}\,\right)\!\cdot {\bf B} \nonumber \\
&=& -\sqrt{2}\,\left(\bar{T}^{\,\gamma}_{\nu;kji'}\,\mu_{i'i} \,B^{\,\gamma}_{ijk} + \bar{B}^{\,\gamma}_{kji'} \,\mu_{i'i} \,T^{\,\gamma}_{\nu;ijk}\right)\epsilon^{\nu\sigma\lambda\rho}F_{\sigma\lambda} v_\rho,
\end{eqnarray}
where $\mu_1$ and $\mu_2$ have the same values as in Eq.\ (\ref{one-body-moments}) except for spin-dependent effects that can be parametrized explicitly. Finally, we obtain the decuplet moment interaction
\begin{equation}
\label{L_muTT}
{\cal L}_{\mu,TT} = \bar{\bf T}_{k'ji}\cdot (2{\bf J\cdot B})\,\mu_{k'k}\,{\bf T}_{ijk} = 3\,\bar{T}_{\alpha;k'ji}\,\mu_{k'k}\,S_\nu \epsilon^{\nu\sigma\lambda\rho} F_{\sigma\lambda} v_\rho\, T^\alpha_{ijk}
\end{equation}
where, again, $\mu_1$ and $\mu_2$ have the same values as in Eq.\ (\ref{one-body-moments}).

\subsection{Two and three-body operators}
\label{subsec:moments_23body}

The usual counting of the chiral invariants for the octet magnetic moments gives nine structures to first order in the quark masses \cite{Bos-L4,Jen-moments}, namely 
\begin{eqnarray}
\label{chiral_terms}
&&{\rm Tr}\,\bar{B}QB, \quad {\rm Tr}\,\bar{B}BQ, \quad {\rm Tr}\,\bar{B}QMB, \quad {\rm Tr}\,\bar{B}BQM, \quad {\rm Tr}\,\bar{B}QBM, \nonumber \\ && {\rm Tr}\,\bar{B}MBQ, \quad {\rm Tr}\,M\,{\rm Tr}\,\bar{B}QB, \quad \quad {\rm Tr}\,M\,{\rm Tr}\,\bar{B}BQ, \quad {\rm Tr}\,MQ\,{\rm Tr}\,\bar{B}B,
\end{eqnarray}
where we have suppressed the factor $\sigma^{\mu\nu}F_{\mu\nu}$ acting on the field $B$. In dealing with the one-body operators, we encountered only the invariants ${\rm Tr}\,\bar{B}QB$ and ${\rm Tr}\,\bar{B}QMB$. We will show here that the remaining seven invariants, and a tenth invariant that distinguishes octet and decuplet moments, arise naturally when we consider two- and three-body operators at the quark level. Our discussion will also suggest the relative importance of the new invariants.

It will be useful to adopt a compressed notation in which we show the structure of the matrix elements in a form that can be used for either the octet or the decuplet. Thus, corresponding to the structure in Eq.\ (\ref{mu_Q}), the basic one-body moment operator will be denoted by
\begin{equation}
{\textstyle\sum_l}\,Q_l\sigma_l
\end{equation}
where $Q_l$ and $\sigma_l\equiv\mbox{\boldmath$\sigma$}_l\!\cdot{\bf B}$ are taken to act on quark $l$ in diagonal or mixed matrix elements between $B$'s and $T$'s with all indices contracted. Any flavor index that is not attached to a $Q$ or $M$ is accompanied by an implied unit matrix $\openone$.  Thus, for the octet,
\begin{equation}
{\textstyle\sum_l}\,Q_l\sigma_l\longrightarrow \bar{B}_{k'j'i'}\,\left(\,Q_{i'i} \sigma_i\,\openone_{j'j}\,\openone_{k'k}+\cdots \right)\,B_{ijk}.
\end{equation}
Upon evaluating the spin matrix element of the $\sigma$'s, we obtain the effective moment operator in Eq.\ (\ref{mu0_BB}). Multiparticle operators are to be evaluated simlarly by applying the the indicated operations right-to-left on the fields in the effective operators, with matrix products assumed when the same flavor index appears in a product such as $M_iQ_i$. 

The invariants we will consider are given in this notation by
\begin{mathletters}
\begin{eqnarray}
\label{invariants}
\label{basic}
&& a.\quad {\textstyle\sum_l}\,Q_l\sigma_l,  \\
\label{QM}
&& b.\quad {\textstyle\sum_l}\,(QM)_l\sigma_l,  \\
\label{mass_insertions}
&& c.\quad (M_i+M_j+M_k)\,{\textstyle\sum_l}\,Q_l\sigma_l + {\textstyle\sum_l}\,Q_l\sigma_l\,(M_i+M_j+M_k),  \\
\label{Thomas_no_M}
&& d.\quad (Q_i+Q_j+Q_k)\,(\sigma_i+\sigma_j+\sigma_k), \\
\label{Thomas_MQ}
&& e.\quad (Q_iM_i+Q_jM_j+Q_kM_k)\,(\sigma_i+\sigma_j+\sigma_k) ,  \\
\label{ThomasQjMi}
&& f.\quad (Q_j+Q_k)\,M_i\sigma_i+M_i\sigma_i\,(Q_j+Q_k) + {\rm permutations},  \\
\label{Thomas_3quark}
&& g.\quad (Q_jM_k+Q_kM_j)\,\sigma_i  + {\rm permutations},  \\
\label{other_mass_spin-spin}
&& h.\quad (M_j+M_k)\,\mbox{\boldmath$\sigma$}_j \!\cdot\! \mbox{\boldmath$\sigma$}_k\,\sigma_iQ_i + {\rm permutations},  \\
\label{Thomas_spin-spin}
&& i.\quad [(Q_j+Q_k)\,(M_j+M_k) + (M_j+M_k)\,(Q_j+Q_k)]\,\mbox{\boldmath$\sigma$}_j \!\cdot\! \mbox{\boldmath$\sigma$}_k\,\sigma_i + {\rm permutations},  \\
\label{symmetric_spin-spin}
&& j\quad \{\mbox{\boldmath$\sigma$}_i \!\cdot\! \mbox{\boldmath$\sigma$}_j + \mbox{\boldmath$\sigma$}_j \!\cdot\! \mbox{\boldmath$\sigma$}_k +\mbox{\boldmath$\sigma$}_k \!\cdot\! \mbox{\boldmath$\sigma$}_i, \mbox{\boldmath$\cdot$}\},   
\end{eqnarray}
\end{mathletters}
where $\mbox{\boldmath$\cdot$}$ in ($j$\/) is any of the invariants ($a$\/)--($g$\/). 

It is useful in interpreting these invariants to recall that the baryon moments obtained in dynamical models such as that of Brambilla {\em et al.} \cite{brambilla} appear as sums of single-particle moments $\mu_i\approx\left<eQ_i/2E_i\right>$ where $E_i$ is the kinetic energy of quark $i$, plus a set of Thomas precession terms \cite{Como}. This is shown in detail in \cite{phuoc-diss} in a quenched approximation to QCD. The existence of the Thomas terms follows in the context of the semirelativistic Hamiltonian in Eq.\ (\ref{H_brambilla}) from the replacement of ${\bf p}_i$ in by ${\bf p}_i-eQ_i{\bf A}$, with ${\bf A}$ the vector potential for the static magnetic field $\bf B$. The diagonal Thomas precession terms are associated with the spin-same-orbit interaction in Eq.\ (\ref{H_brambilla}), and give a multiplicative correction to $\mu_i$. The two-body spin-other-orbit terms introduce new, nonadditive structure, and are important in improving the simple quark-model fits to the moments \cite{DH-loop-moments}. These terms are proportional to the short-distance spin-dependent part of the potential of order $\alpha_s$ divided by $E_iE_j$, so are expected to be small. 

In the presence of symmetry-breaking mass terms, the dynamical matrix elements in the $\mu_i$ differ from those in the symmetrical limit. There is a direct change associated with the change in $1/E_i\approx 1/m_i$ with the effective mass of quark $i$. There are also indirect changes associated with the effects of the quark masses on the baryon wave functions. The Thomas terms are changed similarly, with new two-body components associated with changes in the factor $1/E_iE_j$ with respect to the masses $m_i$ and $m_j$, and a three-body component that reflects the dependence of the matrix element on the mass of the third quark. Finally, all of the preceding matrix elements are modified by the changes in the baryon wave functions caused by the spin-spin interaction in the Hamiltonian. In the equal-mass limit, these corrections, expected to be quite small, distinguish the effective octet and decuplet moments. They contribute further terms with the same structures as above for unequal masses.  With this backgound, we can interpret the contributions to the baryon moments in Eqs.\ (\ref{basic})-(\ref{symmetric_spin-spin}) dynamically.

The invariant $(a)$ gives the baryon moments in the symmetrical limit in the absence of spin-dependent forces as already discussed. The second one-body term $(b)$ corrects for the dependence of the dynamical matrix element on the symmetry-breaking mass of the quark in question, that is, the change of $<1/E_i>$ with $m_i$ in the model. The invariant $(c)$ includes the indirect first-order corrections from the effects of the other quark masses on the matrix element, and begins to introduce a dependence of the effective moment of the quark on its environment. We expect these corrections to be smaller than the direct correction $(b)$.

The term $(d)$ arises from the Thomas terms in the symmetrical limit, and is independent of $M$. It corresponds at the quark level to the two-body interaction
\begin{equation}
(\bar{q}_lQ_{ll'}q_{l'})\,(\bar{q}_p \mbox{\boldmath$\sigma$}\!\cdot {\bf B} q_{p'}).
\end{equation}
We can evaluate matrix elements of this operator directly as in Sec.\ \ref{subsec:meson-baryon_couplings} and convert the results to effective baryon-level operators. Alternatively, a simpler calculation based on Eq.\ (\ref{Thomas_no_M}) and the observation that $\mbox{\boldmath$\sigma$}_i + \mbox{\boldmath$\sigma$}_j + \mbox{\boldmath$\sigma$}_k$ is the total spin operator, so acts only on the spinor index of $B$ or $T$, gives the structure
\begin{equation}
\label{(d)_matrix_element}
\left(\,\bar{B}_{k'ji}\,Q_{k'k}\mbox{\boldmath$\sigma$}\,B_{ijk} + \bar{B}_{kj'i}\,Q_{j'j}\mbox{\boldmath$\sigma$}\,B_{ijk} + \bar{B}_{kji'}\,Q_{i'i}\mbox{\boldmath$\sigma$}\,B_{ijk}\,\right)\!\cdot {\bf B}
\end{equation}
for the effective octet operator. The result for ($e$) has the same structure with $Q\rightarrow QM$, reflecting the corrections to the Thomas terms associated with the mass of the quark that couples directly to the magnetic field.

Converting the expression in Eq.\ (\ref{(d)_matrix_element}) and its analog for $Q\rightarrow QM$ to matrix form, we obtain the pure $F$-type interactions
\begin{equation}
{\rm Tr}\,\bar{B}[Q,B],\quad {\rm Tr}\,\bar{B}[QM,B].
\end{equation}
While $F$-type invariants already appear in the additive or one-body octet quark moments in Eq.\ (\ref{one-body-moments}), they are accompanied by specific $D$-type and double-trace terms. As a result, the pure $F$-type contributions from ($d$) and ($e$) depart from the additive model. The off-diagonal parts of ($d$) and ($e$) are two-body rather than one-body operators, so additivity is lost.

The two-body invariant ($f$\/) and the three-body invariant ($g$\/)  in Eqs.\ (\ref{ThomasQjMi}) and (\ref{Thomas_3quark}) give additional quark-mass corrections to the Thomas terms, ($f$\/) from the direct mass correction for the quark whose spin is involved in the interaction, and ($g$\/) from the indirect effect of the third quark on the matrix elements. ($h$\/) and ($i$\/) describe the effects of the mass-dependent parts of the spin-spin interactions given in Eq.\ (\ref{spin-spin-Dirac}) on the matrix elements for the leading ($a$\/) and ($d$\/) terms, taken to first order in $M$. We have dropped further terms which involve $\{(M_i+M_j)\,\mbox{\boldmath$\sigma$}_i \!\cdot\! \mbox{\boldmath$\sigma$}_j,\,Q_i\sigma_i\}$, $\{(M_i+M_j) \,\mbox{\boldmath$\sigma$}_i \!\cdot\! \mbox{\boldmath$\sigma$}_j, \,(Q_j+Q_k)\sigma_i\}$, and permutations in writing ($h$\/) and ($i$\/). These terms, while present, reduce to the previous structures. Because the leading moment matrix elements involve averages over the entire baryon, while the spin-spin terms are weak and of short range, we expect ($h$\/) and ($i$\/) to be unimportant.

The last invariant, ($j$\/), represents the effect of the symmetrical part of the spin-spin interaction on the moment matrix elements through changes in the baryon wave functions. This invariant induces an overall multiplicative change in the decuplet matrix elements relative to the octet matrix elements, the operator $\mbox{\boldmath$\sigma$}_i \!\cdot\! \mbox{\boldmath$\sigma$}_j + \mbox{\boldmath$\sigma$}_j \!\cdot\! \mbox{\boldmath$\sigma$}_k +\mbox{\boldmath$\sigma$}_k \!\cdot\! \mbox{\boldmath$\sigma$}_i$ having the value $\pm 3$ on decuplet (octet) fields. This affect should again be quite small, and is irrelevant in the absence of precision measurements of the decuplet moments.

It is possible to put all of the octet effective operators in matrix form, but the results are cumbersome and not especially illuminating. We note only that that ($a$\/)-($e$\/) already involve all the standard structures in Eq.\ (\ref{chiral_terms}), but not with completely independent coefficients. That independence is provided by ($f$\/)-($i$\/).

We conclude by noting that the remarkable success of the additive quark madel in describing the octet baryon moments follows directly from the relative weakness of the spin-spin interactions seen in dynamical models. Our results here are independent of the detailed dynamics as far as the structure of the moments in relativistic effective field theory is concerned. However, dynamical information is clearly very useful in anticipating the importance of different chiral structures, and in interpreting those structures in a way that is obscured in the obscured in the usual matrix representations.

\section{CONCLUSIONS}
\label{sec:conclusions}

Our objective in this paper was to demonstrate the advantages of using the three-flavor-index representations $B_{ijk}^{\,\gamma}$ and $T_{ijk}^{\, \mu\gamma}$ for the octet and decuplet baryon fields in HBChPT. We have considered only the leading terms in the momentum expansion and the first-order corrections in the symmetry-breaking quark mass $m_s$. We will extend the analysis to loop corrections in a separate paper, where we will show that their apparently small effect on fits to baryon masses and moments in HBChPT is a consequence of the structure of the theory \cite{phuoc-diss,jaczko-diss}. However, we have already obtained a number of useful results which we think demonstrate the advantages of the method despite its lack of familiarity and the somewhat more complex calculations involved. 

We find, for example, that the SU(6) relations for meson-baryon couplings and baryon masses and moments appear automatically at leading order in the momentum expansion and O($m_s^{\ 0}$\/) in the masses if the effects of spin-spin correlations or interactions are neglected. The SU(6) relations are broken by spin-spin interactions which introduce, among other effects, a decuplet-octet mass splitting. They persist approximately for real baryons because the spin-dependent interactions are short-ranged and perturbatively weak as seen in successful dynamical models \cite{brambilla,isgur,kogut}. The simple connection of the three-flavor-index form of the chiral expansion to the underlying dynamics is the key to this interpretation. It is obscured in the usual matrix representation of the octet fields, in which the approximate validity of SU(6) relations in the chiral expansion appears to be accidental.

We found also that the terms in the new chiral expansion for the baryon masses have a structure identical to that assumed in the NRQM even though we are dealing with a relativistic effective field theory. This correspondence, which holds through O($m_s$\/), is essentially kinematical. We are dealing with the most general description of the spin and flavor correlations in the chiral matrix elements, and this structure is the same as that modeled in the NRQM. The connection goes somewhat further. The internal structure of the baryon is averaged out in matrix elements at leading order in the momentum expansion. The residual quark degrees of freedom move with the heavy, nonrelativistic baryon in HBChPT, hence appear as ``nonrelativistic'' constituent quarks.

In the case of the baryon moments, we understand the striking success of the additive quark model as resulting from the dominance of the one-body operators in our expansion over the nonadditive two- and three-body operators. The latter are again proportional perturbatively to the relatively weak spin-spin interactions. The separation of one- and more-body operators is also the key to understanding the detailed structure of loop corrections in the chiral expansion \cite{phuoc-diss,jaczko-diss,DHJ}. We believe this method should be quite useful in the analysis of HBChPT in more general situations, and provide physical interpretations of the terms which appear through the connection with dynamical models. Some obvious applications include the analyses of the strong baryon-meson couplings beyond the symmetrical limit, of the structure of the weak currents, and of low-energy scateering amplitudes.

\acknowledgments
This work was supported in part by the U.S. Department of Energy under Grant No.\ DE-FG02-95ER40896, and in part by the University of Wisconsin Graduate School with funds granted by the Wisconsin Alumni Reseach Foundation. One of the authors (LD) would like to thank the Aspen Center for Physics for its hospitality while parts of this work were done.

\end{document}